# Spin covalent chemistry of carbon

E.F. Sheka

Institute of Physical Researches and Technology, Peoples' Friendship University of Russia (RUDN University), 117198 Moscow, Russia;
sheka@icp.ac.ru

**Abstract.** This review presents the covalent chemistry of carbon within the spin-radical concept of electron interaction. Using the language of valence bond trimodality, the regions of classical spinless covalence and its spin counterpart are defined. Carbon is the only element exhibiting spin covalent chemistry. Classical covalent chemistry of carbon concerns molecular substances whose valence bond structure includes segregate or chained single $sp^3C - C$ bonds. Substances with double $sp^2C = C$ and triple $sp^1C \equiv C$ bonds are the subject of spin covalent chemistry of carbon. The mathematical apparatus of spin covalence forms the basis of algorithms governing the chemical modification of carbon substances, polymerization processes, and catalysis involving them, making it possible to supplement the empirical spin covalent chemistry of carbon with its virtual analog.

**Introduction**

The covalent realm of modern chemistry is populated by pairs of bonded neutral atoms, representing one of the most deeply ingrained and convenient graphical representations of interatomic interactions in matter. The wide diversity of these pairs, possible for virtually any element in the Mendeleev's periodic table, is the fundamental basis of this representation's perfection. Carbon stands out in this context—an atom that forms only covalent bonds with atoms of other elements and possesses the unique ability to form three types of covalent bonds with itself. The bond trimodality is so widespread in nature that it allows us to distinguish a distinct, extensive class of substances whose atoms are held together by purely carbon covalent $C \leftrightarrow C$ bonds, realized as a three-mode set of $sp^3C - C$, $sp^2C = C$ and $sp^1C \equiv C$ bonds. The prefix $sp^n$ used to describe them denotes one of three possible configurations of the $2s$ and $2p$ electrons that provide the chemical bond between two carbon atoms. The empirical discovery and theoretical justification of the trimodality of $C \leftrightarrow C$ bonds underly the covalent chemistry of pure carbon compounds, which consist of both branched networks of and lone C$_2$ atomic pairs. The first configuration is characteristic of solid-state pure $sp^n$ carbons and polymers, while the second applies to individual molecules, the number of which is virtually countless.

This article discusses $sp^n$ carbons, both solid-state and molecular, whose remarkable properties allow us to uncover the full depth of the chemical nature of this element. The article is structured as follows. Section 2 briefly presents the empirical picture of solid-state and molecular $sp^n$ carbon. The determining factors of the spin character of $sp^n$ carbon chemistry and the spin emergents of its covalent chemistry are presented in Section 3. The spin chemistry of alkane, alkene, and alkyne bonds is described in Sections 4, 5, and 6, respectively. Carbon catalysts are discussed in Section 7 in light on their spin covalence. Section 8 presents aposteriori reflections and concluding remarks.

## 2. Carbon in the Language of Covalent Bonds

The Earth's carbon world is extremely rich and diverse. It comprises most of the planet's mineral kingdom and all living matter. Thus, Nature, by bringing human beings into contact with carbon, immediately places them in a world of big numbers, touching upon virtually every aspect of the carbon environment: many species, many variations of each species, countless cells, a vast number of atoms and molecules. From the moment of contact with the carbon component of the world, humans have ceaselessly searched for the patterns that govern it. The entire development of natural science represents gradual steps toward understanding how this world unfolds. Discovered first patterns, then replaced by others, carefully selected bits of knowledge, and chemical, biological, physical, geological, and cosmogenic understandings of the role and place of carbon in Nature has been gradually developed. Today, the birth of a mathematical view of carbon as the foundation of the carbon world of the metaverse is underway. The carbon atom and the molecules it formed laid the foundation of these concepts. The atom as the basic building block and the molecule as the fundamental configuration of atoms, ensuring the existing reality of large numbers.

Molecular theory gave rise to the concept of interatomic bonds, and from that moment on, the world of carbon took on new colors. As it turned out, Nature endowed the carbon atom with the unique ability to form stable bonds with virtually all chemical elements, particularly highlighting its bonds with itself, noting their distinctive three-modal nature: $sp^1$, $sp^2$, $sp^3$. This trimodality concept proved not only a convenient mathematical device but also touched upon the depths of the multi-dimensional nature of the carbon world, greatly increasing the variability of its chemical, biological, physical, geological, and cosmogenic manifestations, on the one hand, and serving as a fundamental, common concept for all of these manifestations, on the other. The differences in the properties of material carbon, determined by the trimodality of its bonds, are also common to all manifestations, regardless of whether they are observed in space or in a living organism on Earth.

This idea of the conceptual unity of the carbon world underlies the discussion below. The subject of this discussion is digital twins, which represent molecular configurations of carbon atoms corresponding to solid forms of elemental carbon (crystals and amorphous forms) and a selected set of real-life molecular carbon. The trimodality of covalent $C \leftrightarrow C$ bonds and its influence on the properties of digitized material objects is the main focus of this discussion. The patterns established by this approach represent a common property of the entire carbon world.

### 2.1. Crystalline Carbon

Solid carbon is a unique material with an internal molecular structure, which can be viewed most favorably from the perspective of $sp^1$, $sp^2$ and $sp^3$ trimodality. This material carbon occupies a significant portion of the Earth's crust [1] and is also present in small quantities in space. A large part of the material solid carbon is produced industrially. Figure 1 shows the currently known crystal structures of elemental carbon as one of the possible sets of their digital twins. The well-known crystalline diamond with a cubic structure (Figure 1a) opens the presentation. Each carbon atom is bonded to four nearby (symmetrically located at the tetrahedron vertices) covalent $sp^3 C-C$ bonds. An ideal crystal can be imagined as a single giant $sp^3$ molecule [2]. Natural diamond deposits, estimated at several million carats, are widespread across the planet. Compared to this crystalline cubic carbon giant, lonsdaleite, or hexagonal diamond, is an alien from outer space and is found primarily in meteoritic debris [3,4] (Figure 1b). As with diamond, its molecular chains of atoms are formed by $sp^3 C-C$ bonds.

In Figure 1c, the $sp^3$ structure gives way to the $sp^2$ structure of graphite, whose triples of $sp^2 C=C$ bonds form a planar trigonal configuration [5]. Unlike the closely packed $sp^3$ structures, graphite consists of parallel layers formed by regular hexagons of carbon atoms linked to each other by $sp^3 C=C$ bonds. As usually accepted, atoms of adjacent layers are not covalently bound and the layers are coupled by Van der Waals attraction. There are two modifications of graphite: α-graphite (hexagonal) and β-graphite (rhombohedral), which differ in the layer arrangement. β-graphite is not observed in pure form, as it is a metastable phase. However, in natural graphite, the rhombohedral phase content can reach 30%. Graphite is widespread in nature, with estimated reserves of 250-300 million tons.

As can be seen from the figure, diamond is a three-dimensional structural carbon modification composed of $sp^3$ atoms. Graphite is a two-dimensional layered allotropic carbon modification formed by $sp^2$-hybridized atoms. The third one-dimensional linear allotropic modification, corresponding to the $sp^1$ hybridization, could not be discovered for a long time. This substance was first obtained artificially by Soviet chemists in the early 1960s and was named "carbyne" [6]. According to the discoverers, the most difficult thing for them was to determine what kind of bonds connected the carbon atoms in carbyne into a chain. These could be alternating single and triple bonds ($-C \equiv C - C \equiv C-$) of polyyne chains, only double bonds ($= C = C = C = C =$) of polyene chains, or both simultaneously. Over time, it was possible to prove that there are no double bonds in the carbyne. This substance, consisting of polyyne chains, was named α-carbyne. According to microdiffraction data, single-crystal films of α-carbyne possess a hexagonal lattice [7,8]. Figure 1d shows a carbyne spatial model. The numerals (I-5) denote carbyne chains, the positions of which in a unit cell of the carbyne crystal in a (0001) projection is shown in the inset. It can be seen that the carbyne lattice is double-layered. The lower layer consists of closely packed carbyne chains. The columns of the chains in the upper layer have vacancies presumably due to impurity atoms such as Fe or K intercalated between the layers. These atoms can provide saturation of the dangling bonds.

In 1968, carbyne-like carbon was found in geological rocks formed in a meteorite crater [9]. To date, approximately two tens of carbyne-like materials differing in structural parameters have been artificially synthesized and revealed in nature [10–13]. Strong chemical activity and extreme instability at ambient conditions characterize carbyne, an infinite $sp^1$ - $sp^3$ carbon chain. As a result, much less has been explored about carbyne as compared to other carbon allotropes. Although end-capping groups can be used to stabilize carbon chains, length limitations are still a barrier for production. To overcome the difficulty, a method for the bulk production of long carbyne linear chains protected by thin double-walled carbon nanotubes and composed of more than 6000 atoms has been proposed [14].

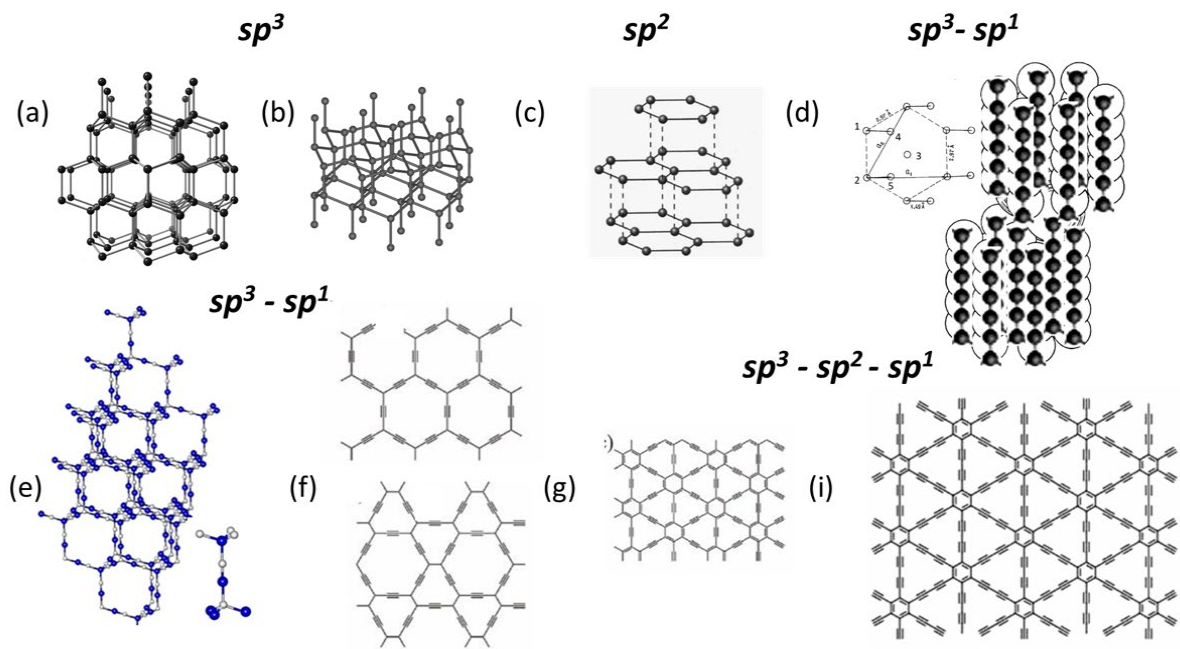

Figure 1. Digital twins of crystalline carbon allotropes. (a) Diamond. (b) Lonsdaleite. (c) Graphite. (d) Carbyne. (f) Y-diamond. (e) α (top) and β (bottom) Grahynes. (g) γ Grahyne. (i) Graphdiyne. Images taken from open sources.

The existence of crystalline carbyne revealed a very important property concerning the covalent bonds of carbon. It had to be recognized that while stable polyyne configurations of carbon atoms linked solely by $sp^1 C \equiv C$ bonds are impossible, a stable configuration can nevertheless be created by allowing these bonds to coexist with $sp^3 C - C$ and $sp^2 C = C$ bonds. As it turned out, in addition to carbyne, it is possible to construct regular carbon configurations involving a variety of bond types. Thus, preliminary hints at the carbon $sp^3 - sp^1$ 3D structures (which were called *yne*-diamonds or Y-diamonds) can be found in early works [14–16], where the insertion of acetylenic ($-C \equiv C -$) or diacetylenic ($-C \equiv C - C \equiv C$) linkages inside the $sp^3$ skeleton of diamond was assumed. Figure 1e presents the $sp^3 - sp^1$ configuration, demonstrating the structure of a virtual Y-diamond crystal [17, 18]. This mixed-bond configuration is predicted to be stable, so Y-diamond crystals may be synthesized in the near future.

As can be seen from Figures 1a and 1e, Y-diamond is the result of a structural transformation of diamond, in which all $sp^3 C - C$ bonds are replaced by ternary $sp^3 - sp^1 - sp^3$ chains. A similar transformation, but affecting all $sp^2 C = C$ bonds of the original two-dimensional $sp^2$ carbon structures, leads to the formation of stable α and β graphynes [19], the digital twins of which are shown in Figure 1f. It also turned out that to obtain a stable planar regular structure, it is sufficient to replace one third of the $sp^2 C = C$ bonds of graphite with acetylenic ($-C \equiv C -$) or diacetylenic ($-C \equiv C - C \equiv C$) linkages [19-21]. Such a transformation in the first case leads to the regular structure of γ graphyne (see Figure 1g), and in the second - to graphdiyne (Figure 1i). Calculations showed the stability of structures of both types, as well as other graph-n-ynes (n = 3, 4, 5, etc.). Numerous attempts to synthesize this group of substances have been successful only in relation to graphdiyne, the crystalline films of which have become the subject of careful study [22-26].

A brief look at carbon crystallography, presented above, shows that $sp^3$ and $sp^2$ bond configurations are favorable for the formation of stable, regular structures, supporting the existence of millions of tons of diamond and graphite deposits on Earth and the existence of crystalline carbon in space. As for $sp^1$ bonds, they cannot form regular structures on their own, so the latter are formed only by combining $sp^1$ bonds with $sp^3$ and $sp^2$ ones.

## 2.2. Amorphous Carbon

The trimodality of valence electron configuration of carbon atoms forms the foundation of the unique three-mode amorphous state of carbon solid. From the fundamentals of solid-state physics, $sp^3$, $sp^2$, and $sp^1$ amorphous carbons ($aC$s) are different species characterized by conceptually different short-range orders, namely, groups of tetrahedrally bonded $sp^3$ configured atoms, as well as nanoscale-size-restricted $sp^2$ graphene domains, and $sp^3$- $sp^1$ carbyne chains. Figure 2 provides a basic overview of the current state of this carbon understanding.

As in the case of crystals, $aC$ has an internal molecular structure, the configuration of which depends on the type of $C \leftrightarrow C$ bonds involved. Thus, tetrahedral $aC$ ($ta - C$) in Figure 2a [27, 28] represents the well-known configuration of atoms linked by $sp^3 C-C$ bonds. Another tetrahedral amorphous allotrope, $Q$-carbon in Figure 2b [29–31], is a densely packed metastable phase formed by ultrafast quenching of carbon melt in a super-undercooled state. After quenching, diamond tetrahedra are randomly packed with >80% packing efficiency. Both $aC$s are not natural substances and constitute only a small part of the vast amorphous wealth of carbon. The central part of the latter is occupied by the $sp^2$ $aC$. Recent purposeful studies [32-37] have completed the gradually nascent change in the view of the solid as a well-known, familiar physicochemical subject, the beginning of which was laid back in 1941 [38], transferring it to the rank of high-tech material of the modern nanotechnology. We are talking about material known to humankind since the first decomposed bonfire, which left behind black soot and charcoal. Today, this applies to billions (anthracite coal) - millions (shungite carbon) - thousands (anthraxolite and blackcarbon coatings, which are present everywhere and accompanying almost all minerals in nature) tons of only discovered natural deposits and hundred-million tons of synthetic black carbon produced industrially. All this black carbon wealth is $sp^2 aC$, or a monoatomic solids without long-range order, the atoms of which form $sp^2 C = C$ configured valence bonds with each other. It has a unique common basis, namely, nano-micro-scale molecular compositions of hexagonal honeycomb structures of carbon atoms (graphene domains) framed by various heteroatom necklaces composed of oxygen, hydrogen, nitrogen, sulfur, halogen and so forth. These necklaced graphene molecules are basic structural units (BSUs), varying in size and shape, as well as differing in the necklace chemical composition depending on the history of origin and/or method of production of the black carbon, and form the first level structure of the bodies. The discovered and experimentally confirmed graphene nature of this black gold has led to a revolutionary revision of the theory, modeling and interpretation of the experiments related to this class of solids [37].

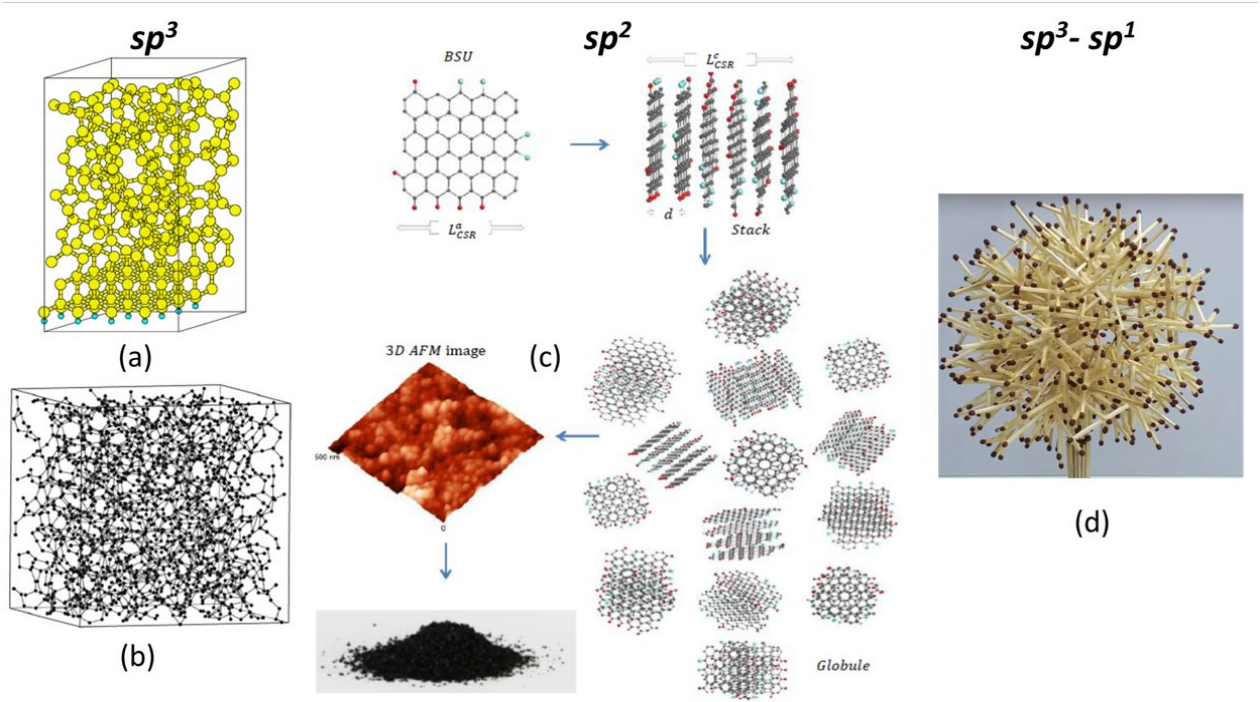

**Figure 2.** Digital twins of amorphous allotropes of carbon. (a) Tetrahedral amorphous carbon $ta-C$. (b) $Q$-carbon. (c) $sp^2$ graphene-carbon. (d) $sp^3$-$sp^1$ carbyne-carbon. Images taken from open sources.

The second level structure pf the $sp^2 aCs$ is provided with nano-thick BSUs stacks, which are confidently recorded by X-ray and neutron diffraction structural studies of $sp^2\ aCs$ of all types [33]. The third-level structure of these amorphics reliably follows from the porous structure evidently observed experimentally [39,40]. It is constructed from the BSUs stacks but the final compositions depend on the stacks' lateral dimension. When the latter is at the first nanometer level, the composition presents globules of ~10 nm in size, which corresponds to pores, size of which is first nanometers as well. Further aggregation of globules leads to the formation of micro-nanosize agglomerates with pores of tens-to-hundreds nm. Such a structure is typical to natural $aC$s such as shungite carbon, anthraxolite, anthracite as well as black carbon coating of diamonds [41], mixed carbon-silica spherical 'sweets' [42], black carbon in meteorites [43,44] and none of the exclusions has been known so far. Figure 2c schematically presents the evolution of this type of amorphic structure from a single BSU to macroscopic powder. In contrast to natural bodies, synthetic $sp^2$ aCs are characterized by a large dispersion of BSUs ranging in size from units to tens and/or over hundreds of nanometers. At the low-limit end of the dispersion, the amorphic structure is similar to that of natural species described above. At the high-limit end, the BSU size does not prevent BSUs from packing in nanosize-thick stacks, while the latter are extended and further packed into a paper-like structure.

In contrast to $sp^3$ and $sp^2\ aC$s, information on $sp^1\ aC$s is currently practically absent, which indicates great difficulties due to both the existence of carbyne chains of limited length and the complexity of their mutual packing. We touched on this problem above when describing the crystallization of carbyne chains. Since the end atoms of the chains are very reactive, chemical approaches tend to stabilize the chains by end groups. Different end groups have been used, ranging from hydrogen or metal atoms to larger complexes. With larger end groups, chains of up to 44 atoms have been synthesized [45]. It is assumed that the extended end groups prevent cross-linking of the chains by keeping them at a distance. Calculations show that both regular and irregular packing of carbyne units is possible. The successful implementation of the former is described in the previous section. Irregular, or amorphous, packing can be imagined as a child's 'hedgehog made of matches', shown in Figure 2d. And only the future will tell how close our idea is to reality.

Concluding the above comparative analysis of the structures of solid carbon allotropes, it should be recognized that the trimodality of the covalent bonds between carbon atoms is the primary reason for their wide diversity. It is easy to imagine that similar vivid images would accompany the description of any characteristic property of carbon molecules. This, in turn, allows us to say that the type of covalent bond for each carbon atom is a determining factor in its subsequent fate and behavior within a team of surrounding atoms. This means that the analysis of these bonds becomes the cornerstone of the entire carbon universe, from its chemistry to its biology and cosmology. This fundamental problem will be further explored below within the framework of the spin theory of the carbon atom covalence.

## 3. Lengths of $C \leftrightarrow C$ Bonds as a Governing Factor of the Carbon Covalent Chemistry

### 3.1. Spin-Radical Concept of the Carbon Atoms' Covalence

Stretching and breaking of chemical bonds, leading to open-shell character of molecular electronic systems, present key points of the spin-radical concept of the covalence. The main statement declares that open-shell character of both covalent carbon solids and molecules is the main mechanism responsible for peculiar properties of the species at microscopic level. Basically, the open-shell character is provided with either initially unpaired odd electrons, which do not participate in the formation of covalent bonds, or effectively unpaired electrons that are withdrawn from the covalent bonding, in whole or in part, due to stretching of interatomic distance. In both cases spacing between the electrons must exceed a critical value $R_{crit}$ characteristic for partners of the covalent bond under consideration.

Theoretically, the bond concept has come a long way of development alongside the electron theory of chemical matter, and its development is still ongoing. Particular epochs are associated with the valence bond theory [46], molecular orbital theory [47], and density functional theory [48]. A comprehensive collection of reviews exhibiting the modern concepts of the chemical bonding is presented in two-part edited collections [49,50]. These theoretical approaches have laid the foundation of quantum chemistry aimed at obtaining equilibrium multi-atomic configurations. However, a direct solution of Schroedinger's equation does not point to the bond within a particular pair of atoms. Computationally, the bond justification consists in finding evidence in space related to the electron density distribution in the frame of either Bader's atom-in-molecules theory [51] or some of its developments (see Refs. [52,53] and references therein). Empirically, in the majority of cases, the bond between two atoms is justified by comparing the interatomic distance with one of standard bond lengths accumulated on the basis of numerous structural data. In view of this interrelation, on practice, the chemical bond is mainly associated with this structural identificatory, with respect to which one can speak about 'bond forming', 'bond stretching', or 'bond breaking'. Speaking about the length of a covalent bond, one usually addresses the data tabulated in numerous tables and presented in numerous handbooks (see, for example, Refs. 54,55). As seen from the data, bond lengths for the same pair of atoms in various molecules are rather consistent which makes it possible to speak about standard values related to particular pairs of atoms. A standard length of 1.09Å is attributed to the C-H pair while the lengths of 1.54 Å, 1.34 Å, and 1.20Å are related to single, double and triple $C \leftrightarrow C$ bonds, respectively. Complicated as a whole, the set of the available data on bond lengths and bond energies provides a comprehensive view on the equilibrium state of molecules and solids. On the background of this self-consistency, the detection of extremely long bonds, such as single $sp^3C - C$ bonds of 1.647 Å, 1.659 Å, and 1.704 Å instead of 1.54 Å [56] and $sp^3C - O$ bonds of 1.54 Å [57] and 1.622 Å [58] instead of 1.43 Å not only looks as a chemical curiosity but raises the question of the bounds of covalent bonding. Two other questions are closely related to the matter: 1) to which extent a chemical bond can be stretched and 2) on which length its breaking occurs. Empirically, this usually concerns subjectively made estimations of critical values of a possible elongation of bonds that broadly varied. Thus, the width of the region of admissible values of bond's lengths is significantly varied in different computer programs aimed at molecular structure imaging. As for a bond rupture, this problem is the most uncertain and the rupture is considered as a final result of a continuous stretching only.

The problem of theoretical justification of the chemical bond stretching and breaking concerns the criteria according to which the considered bond is still alive or ceases to exist. Until now, two approaches have been usually exploited. The first, based on the atom-in-molecules theory [51], concerns the bond critical point within the electron density distribution over an atomic composition, evidence of which is considered as a proof of the bond existence. However, as shown [58], the criterion, computationally realized, is not reliable in the case of weak coupling due to which it cannot be used to fix the bond breaking. The second approach overcomes the difficulty addressing directly to the correlation of electrons involved in the bond [59] addressing the entanglement among any pair of orbitals. The obtained results showed that electron correlation is indeed the main determinant of stretching and breaking of chemical bonds and the quantitative measure of the correlation may serve as criteria for the fixation of the above processes.

Unrestricted two-determinant Hartree-Fock (UHF) formalism is quite suitable for a quantitative description of electron correlation thus providing four criteria able to characterize the extent of the event [60]: (i) misalignment of the energy of RHF ($E^R$) and UHF ($E^U$) solutions $\Delta E^{RU}$, where $\Delta E^{RU} = E^R - E^U$; (ii) spin expressed contamination via misalignment of squared spin $\Delta \hat{S}^2$; here $\Delta \hat{S}^2 = \hat{S}^2_U - S(S+1)$, $\hat{S}^2_U$ is the UHF squared spin while $S(S+1)$ presents the exact value; (iii) appearance of effectively unpaired electrons of $N_D$ total number: (iv) molecular magnetism governed by exchange integral $J = \dfrac{E^{UHF}_{S=0} - E^{UHF}_{S=S_{max}}}{S^2_{max}}$. Moreover, the HF level of the theory is quite appropriate for understanding the basic aspects of bonding [61].

### 3.2. Spin Emergents of Carbon Covalent Bonds

Besides UHF emergents $\Delta E^{RU}$, $\Delta \hat{S}^2$, $N_D$, and $J$, completely describing the open-shell nature of the molecular configuration under consideration, dependent on the interatomic distance $R$ $N_D(R)$ graphs and their singularities present a perfect benchmark for a quantitative description of stretching and breaking of chemical bonds [62]. Bonds formed by two carbon atoms are the richest in content, and its general representation in the $C \leftrightarrow C$ form covers a set of traditionally matched single $sp^3 C - C$, double $sp^2 C = C$, and triple $sp^1 C \equiv C$ bonds. Throughout the article, the values of the above parameters were obtained using the CLUSTER-Z1 software [63,64] implementing the AM1 version of the semi-empirical unrestricted two-determinant Hartree-Fock (UHF) approach [65]. The program showed itself highly efficient concerning open-shell electronic systems such as fullerenes [66-68], graphene molecules [69], and stable radicals [70].

The $N_D(R)$ graphs in Fig. 3a present a general view on the $C \leftrightarrow C$ bond family on an example of the gradual virtual dissociation of ethane (H₃CC-CCH₃), ethylene (H₂CC=CCH₂), and propyne (H₃CC≡CCH) molecules [62,69]. As seen in the figure, all the studied graphs are of $S$-like shape but significantly different. Thus, the single-bond graph is of one-step $S$-shape while for double and triple bonds $S$-like curves are evidently of two- and three-step, respectively. The number of steps evidently corresponds to the number of electron pairs involved in the relevant C↔C bond. Each of the graphs starts by a horizontal line corresponding to $N_D = 0$. Marked with large red dots, corresponds to the equilibrium length of the bonds $R_{eq}$ while the right-hand bound indicates at which interatomic distance the covalent bonding is violated thus pointing to the largest classical covalent bond length $R_{cov}$, from which the bond stops to be classical and becomes to work as spin-dependent. This region can be characterized by both absolute and relative width $\Delta W_{cov} = R_{cov} - R_{eq}$ and $\delta W_{cov} = \Delta W_{cov}/R_{eq}$ and distinguishes areas where the bonds correspond to the close-shell character of the relevant molecule electronic states. Superscripts $sg, db$, and $tr$ differentiate members of the $C \leftrightarrow C$ family.

When reaching $R_{cov}$, each of the three $C \leftrightarrow C$ graphs undergoes a jump that indicates the beginning of the bond radicalization when stretching. The radicalization gradually proceeds while the interatomic distance increases, although quite differently for the three bonds. Thus, the radicalization of the $sp^3 C - C$ bond of ethane, started at $R_{cov}^{sg} = 2{,}11$Å, is fully completed at $R_{rad} \leq 3$Å and two single radicals are formed. Radicalization of the $sp^2 C = C$ bond of ethylene starts at $R_{cov}^{db} = 1.395$Å and is saturated at the same $R_{rad}$ as for the single bond where a pair of two-fold radicals is formed. However, on the way to a completed radicalization a clearly seen kink on the $N_D(R)$ graph occurs. The kink critical point corresponds to $N_D$ approaching 2 e and, exhibited by differentiating, is located at $R_{k_1}^{db}$ = 2.12Å that is well consistent with $R_{cov}^{sg}$ of the single bond. Therefore, the bond radicalization occurs in two steps, first of which is completed for a pair of π electrons by reaching $N_D \approx 2$ e while the second should be attributed to the separation of the remaining σ electrons until $N_D \approx 4$ e is reached.

$N_D(R)$ graph of the $sp^1 C \equiv C$ bond of propyne, preserving a general $S$-like pattern, shows a two-kink behavior. As seen in Figure 3a, the bond radicalization starts at $R_{cov}^{tr} = 1.24$Å and the first kink is located in the region of $N_D \approx 2$ e at $R_{k_1}^{tr} = 1.40$ Å that is consistent with $R_{cov}^{db} = 1.395$ Å of the double bond of ethylene. In the region of $N_D \approx 4$ e, the second kink is observed, whose critical point at $R_{k_2}^{tr} = 2.10$ Å is consistent with $R_{cov}^{sg}$ of the single bond of ethane. A pair of three-fold radicals at $R_{rad} \geq 3$Å completes the bond breaking. Therefore, a gradual stretching of the $sp^1 C \equiv C$ bond of propyne can be presented as a consequent completed radicalization of two pairs of π electrons first and then terminated by the radicalization of σ electrons followed with the total bond breaking.

Data presented in Fig. 3a allow speaking about a new aspect of chemical bonds concerning their radicalization. It should be remained that the radicalization is just a 'chemical' manifestation of the correlation of bond-involved valence electrons. From this viewpoint, single, double, and triple bonds are drastically different. Thus, the single bond is radicalized in the vicinity of its breaking. Derivation of the $N_D(R)$ graph reliably highlights $R_{cov}^{sg}$ as a clear singularity thus allowing its attribution to the fixation of the bond breaking. In the case of double bond, $R_{cov}^{db}$ determines the parting of π bond while $R_{k_1}^{db}$, which coincides with $R_{cov}^{sg}$, fixes the bond breaking. Similarly, $R_{cov}^{tr}$ on the $N_D(R)$ graph marks the separation of π electrons of the first pair while $R_{k_1}^{tr}$ and $R_{k_2}^{tr}$ manifest the parting of π electrons of the second pair and the remained σ electrons, respectively, just completing the bond breaking. Therefore, according to the observed consistency of $R_{k_1}^{db}$ of ethylene and $R_{k_2}^{tr}$ of propyne with $R_{cov}^{sg}$ of ethane, all the three values can be attributed to the interatomic distance at which any of the bonds of the discussed $C \leftrightarrow C$ set can be considered as broken.

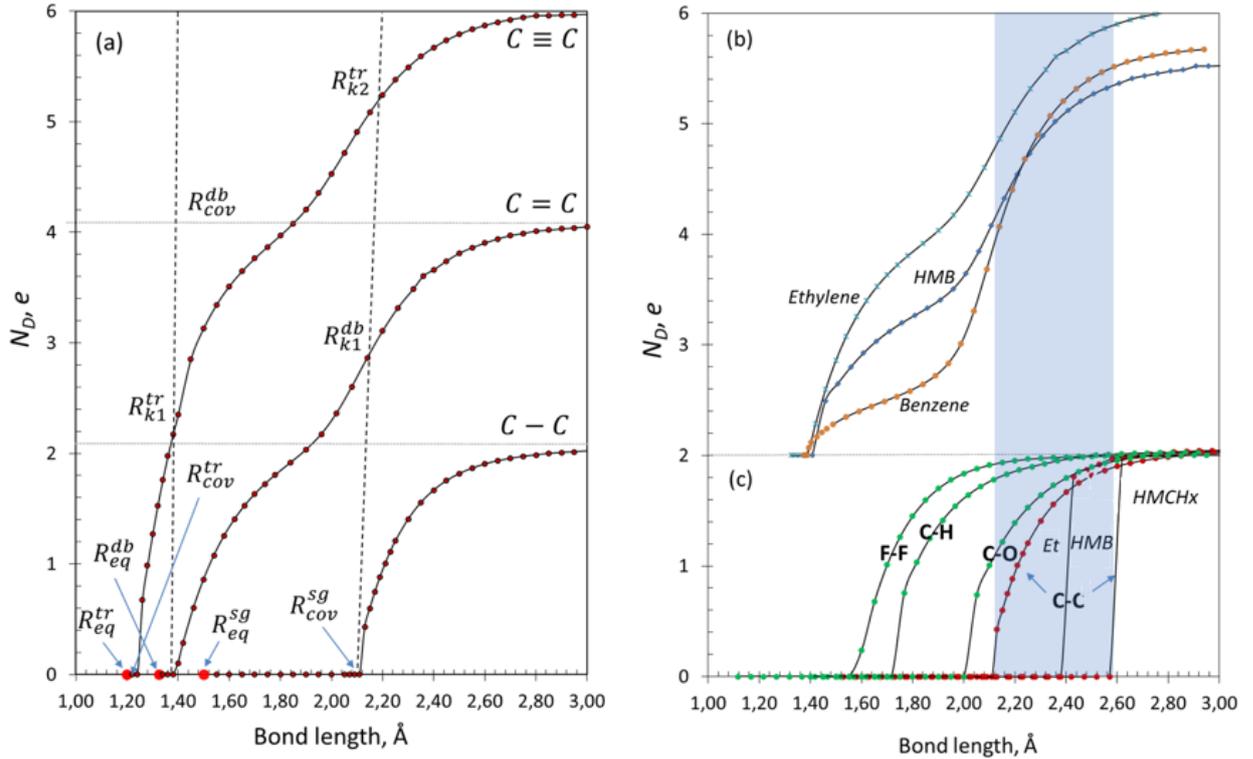

**Figure 3**. $N_D(R)$ graphs of carbon covalent bonds. (a) $sp^3 C - C$ bond of ethane; $sp^2 C = C$ bond of ethylene; $sp^1 C \equiv C$ bond of propyne. Dotted vertical lines mark positions of $R_{cov}^{db}$ and $R_{cov}^{sg}$. Large red points mark positions $R_{cov}$ in all the cases. (b) $sp^2 C = C$ bonds of ethylene, benzene. and hexamethylbenzene (HMB). (c) $sp^3 C - C$ bonds of ethane, cyclohexane (CHx), and hexamethylcyclohexane (HMCHx). Light blue band marks dispersion of the $R_{cov}^{sg}$ depending on the bond structure-elemental surrounding. UHF AM1 calculations.

Fixation of the bond breaking allows introducing such characteristic quantities as the absolute and relative width of the radicalization region $\Delta W_{rad}$ and $\delta W_{rad}$, respectively, that in the case of double and triple bonds of ethylene and propyne are of the form

$$\Delta W_{rad}^{db} = R_{k_1}^{db} - R_{cov}^{db}; \quad \delta W_{rad}^{db} = \Delta W_{rad}^{db} / R_{eq}^{db} \tag{1}$$

for $sp^2 C = C$ bond of ethylene; and

$$\Delta W_{rad}^{tr} = R_{k_2}^{tr} - R_{cov}^{tr}; \quad \delta W_{rad}^{tr} = \Delta W_{rad}^{tr} / R_{eq}^{tr} \tag{2}$$

for $sp^1 C \equiv C$ bond of propyne. The corresponding sets of $R_{eq}, R_{cov}, R_{k_1}, R_{k_2}, \Delta W_{cov}, \delta W_{cov}, \Delta W_{rad}, \delta W_{rad}$ data are listed in Table 1. As seen from the table, while $\Delta W_{cov}$ decreases when going from single to triple bond, $\Delta W_{rad}$ inversely increases. The feature is the main reason for a drastic difference in the chemical activity of the bonds (material bodies and molecules) of different multiplicity.

**Table 1.** Characteristic interatomic distances related to selected covalent chemical bonds [1], Å (data are taken from [69])

| Bonds | Molecules | $R_{eq}$ | $R_{cov}$ | $\Delta W_{cov}$ | $\delta W_{cov}$% | $R_{k_1}$ | $R_{k_2}$ | $\Delta W_{rad}$ | $\delta W_{rad}$% |
|---|---|---|---|---|---|---|---|---|---|
| $sp^3C - C$ | ethane | 1.503 | **2.110** | 0.607 | 40.4 | | | | |
| | cyclohexane | 1.461 | **2.570** | 1.109 | 75.9- | | | | |
| | | 1.485 | | 1.085 | 73.1 | | | | |
| | hexamethyl- | 1.515 | **2.380** | 0.865 | 57.1- | | | | |
| | cyclohexane | 1.532 | | 0.848 | 55.3 | | | | |
| $sp^2C = C$ | ethylene | 1.326 | 1.388 | 0.062 | 4.7 | **2.140** | | 0.752 | 56.7 |
| | benzene [2] | 1.395 | 1.395 | 0.005 [2] | 0 | **2.139** | | 0.744 | 53.3 |
| | hexamethylbenzene [2] | 1.395 | 1.408 | 0.013 [2] | 0.9 | **2.158** | | 0.740 | 53.0 |
| $sp^1C \equiv C$ | propyne | 1.197 | 1.240 | 0.043 | 3.6 | 1.450 | **2.100** | 0.860 | 71.8 |
| $sp^3C - H$ | ethane | 1.117 | **1.717** | 0.600 | 53.7 | | | | |
| | ethylene | 1.098 | | | | | | | |
| | propyne | 1.059 | | | | | | | |
| $sp^3C - O$ | ethyleneglycol | 1.412 | **2.000** | 0.588 | 41.6 | | | | |
| $sp^3Si - Si$ | disilane | 2.418 | **2.950** | 0.532 | 22.0 | | | | |
| $sp^2Si = Si$ | disilene | 2.293 | 1.80 | | | **2.800** | | | |
| $sp^1Si \equiv Si$ | disilyne | 2.31 | 1.64 | | | - | **2.650** | | |
| $sp^3Si - H$ | disilane | 1.466 | | | | | | | |
| | disilene | 1.457 | | | | | | | |

[1] Bold figures correspond to interatomic distances at with the relevant covalent bonds are broken.
[2] The marked molecules belong to a border-regime type determined by small $W_{cov}$ values.

From the above it follows that any chemical bond should be described by a set of characteristics, only one of which, namely, the equilibrium length of chemical bond $R_{eq}$ can be standardized. However, empirical data show that $R_{eq}$ is characterized with a significant dispersion indicating the dependence of the quantity on surrounding atoms. From this viewpoint, the data presented for the considered three molecules may change when going to other atomic composition. Actually, data presented in Figure 3c for single $sp^3C - C$ bonds show that the absolute values of $R_{eq}^{sg}, R_{cov}^{sg}, \Delta W_{cov}^{sg}, \delta W_{cov}^{sg}$ (see Table 1) are different while the qualitative character of the relevant $N_D(R)$ graphs is conserved. Particularly, it should be noted that in polyatomic molecules the radicalization and breaking of these bonds become more abrupt thus significantly narrowing the smoothing of the region of their radicalization.

Oxides and hydrides are the most popular species of the carbon chemistry that is why $sp^3C - O$ and $sp^3C - H$ bonds deserve a particular attention. The relevant $N_D(R)$ graphs presented in Figure 3c are related to the dissociation of single $C - O$ bonds in ethylene glycol (-C-O) as well as $C - H$ bonds of ethane. $R_{eq}^{sg}, R_{cov}^{sg}, \Delta W_{cov}^{sg}, \delta W_{cov}^{sg}$ parameters of the bond are listed in Table 1. As seen in the figure and follows from the table, the bonds one-step behavior is similar to that of $sp^3C - C$ one. The elongation stage $\delta W_{cov}$ constitutes ~40-50%, while the radicalization smoothing of $R_{cov}^{sg}$ is small enough. As in the case of $sp^3C - C$ bonds, one should expect a slight difference in the characteristics of $sp^3C - O$ and $sp^3C - H$ bonds depending on the atomic surrounding.

Double $sp^2C = C$ bonds, similarly to the ethylene one, are characteristic for a large family of alkenes. However, such bonds are more often associated with benzene-based and other aromatic molecules. Their stretching is of extreme significance for a large class of $sp^2$ nanocarbons [71]. Figure 3b presents a comparative view on the dissociation of a single $sp^2C=C$ bond of ethylene and one of the bonds of benzene and hexamethylbenzene. The comparison reveals a common character of the relevant $N_D(R)$ graphs with some difference of $R_{eq}^{db}, R_{cov}^{db}, R_{k_1}^{db}, \Delta W_{cov}^{db}, \delta W_{cov}^{db}, \Delta W_{rad}^{db}, \delta W_{rad}^{db}$ values (see Table 1) as well as a remarkable difference in the graphs' shape. Nevertheless, all the graphs are two-step $S$-like with a kink located in the region of $N_D \sim 2e$. The kink critical points are well consistent with $R_{cov}^{sg}$ of the relevant single $sp^3C-C$ bonds.

Evidently, all the said above can be attributed to species with $sp^1C \equiv C$ bonds. The relevant critical point $R_{cov}^{tr} = R_{crit}^{tr} \cong 1.240$ Å determines the onset of the transformation of molecules involving triple bonds from closed-shell to open-shell ones.

### 3.3. Spin Emergents of Silicon Covalent Bonds

The current around-graphene science has represented a new milestone of activity in the discussion of similarity-and/or-unlikeness of different members of $C \leftrightarrow C$ and higher tetrels $X \leftrightarrow X$ (X=Si, Ge, Sn) families of covalent bonds, which was a hot topic over a century [72]. Now this branch is full of suggestions concerning new prototypes of graphene, foremost of which are based on the equivalence of valence electrons of all tetrels atoms and expected hexagon patterned one-atom-thick planar structures such as silicene, germanene, and tinene-stanene. Covalent radii of tetrels make a series 0.76-0.73-0.69 Å, 1.11 Å, 1.20 Å, and 1.39 Å for carbon ($sp^3$-$sp^2$-$sp^1$), silicon, germanium, and tin, respectively [73]. To form a reliable platform for a comparative analysis, the data presented below are related to molecules of the common structure, namely, ditetralanes $X_2H_6$, ditetrelenes $X_2H_4$, and ditetrylynes $X_2H_2$ ($C_2H(CH_3)$ in the case of carbon) [74,75]. Figure 4a presents $N_D(R)$ graphs related to complete sets of $Si \leftrightarrow Si$ bonds while Figure 4b exhibits the difference between carbon and silicon bonds [62, 69, 74,75]. As seen in the figure, the graphs of silicon covalent bonds behave quite similar to those shown for carbon whilst shifted to longer interatomic distances. The graph of disilane is one-step with clearly seen points $R_{cov}^{sg}$. The graph of disilene demonstrates only the second part of the two-step radicalization, which is a reality for ethylene. The equilibrium interatomic distance $R_{eq}^{db}$ ≈2.3Å for the latter greatly exceeds $R_{cov}^{db}$ at 1.8 Å. Consequently, in contrast to covalently saturated ethylene, equilibrium disilene is almost two-fold radical. When proceeding with the bond elongation, the graph reveals kink positioned at $R_{k_1}^{db}$, well consistent with $R_{cov}^{sg}$. Thus, equilibrium disilene with separated π electrons, continues their dissociation until parting the remained σ electrons at $R_{k_1}^{db} \approx R_{cov}^{sg}$. The three-step radicalized $sp^1C \equiv C$ bond of propyne in Figure 3a is not reproduced as $sp^1Si \equiv Si$ bond of disilyne in Figure 4a. The equilibrium distance $R_{eq}^{tr}$ ≈2.3 Å is positioned much over $R_{cov}^{tr}$ thus presenting almost fourfold radical, which means that π electrons of both pairs are disunited. All parameters of the discussed silicon bonds are listed in Table 1.

Therefore, in contrast to stable carbonaceous materials, possessing $sp^2C=C$ and $sp^1C \equiv C$ bonds, siliceous ones with the shortest $sp^2Si=Si$ and $sp^1Si \equiv Si$ bonds are radicals, chemical activity of which drastically surpasses that of carbon species. The interaction of two odd electrons, which are formed under transformation at any interatomic bond, depends on the corresponding distance $R_{int}$ that is ~1.5 times larger for $Si \leftrightarrow Si$ distances with respect to $C \leftrightarrow C$ ones due to larger size of the atoms. At the same time, the distance $R_{int} = 1.4$ Å is critical for these electrons to be covalently coupled. Above the distance the electrons become effectively unpaired, therewith the more the larger the distance. $sp^2C=C$ bond length of benzene just coincides with the limit that provides a complete covalent bonding of the electrons, transforming them into widely known π electrons. Apparently, this is the way Nature has enabled benzene to play a particular role for establishing and proving the aromaticity concept in the framework of classical covalent chemistry as well as for introducing π electrons in organic chemistry.

Beside the discussed $sp^2C = C$ and $sp^2Si = Si$ bonds, one more double bond is formed involving both carbon and silicon atoms. Figure 4b presents a comparative view on $N_D(R)$ graphs related to the dissociation of ethylene, silaethene, disilene as well as ethane and disilane molecules. Comparing the graphs for $sp^2C = C$ and $sp^2Si = Si$ bonds, a drastic difference becomes evident. In the case of ethylene, one can distinctly see that π electrons govern the molecule continuous dissociation when interatomic distance changes from 1.4 Å to 2.1 Å and then comes the turn of σ electrons until the dissociation is completed at 2.8 Å. At equilibrium the molecule is closed-shell one with the $sp^2C = C$ bond of 1.326 Å in length. Oppositely, π electrons are practically unobservable under disilene dissociation since already in equilibrium they are almost fully transformed into a pair of effectively unpaired electrons ($N_D$=1.78 e). Therefore, disilene has no closed-shell phase at all and in the equilibrium is open-shell one. Essentially different is the situation with the $sp^2Si = C$ bond of silaethene. As seen in the figure, at $R_{eq}^{db}$ = 1.605 Å the molecule is just at the beginning of its open-shell status, radicalization of which constitutes $N_D$ = 0.153 e.

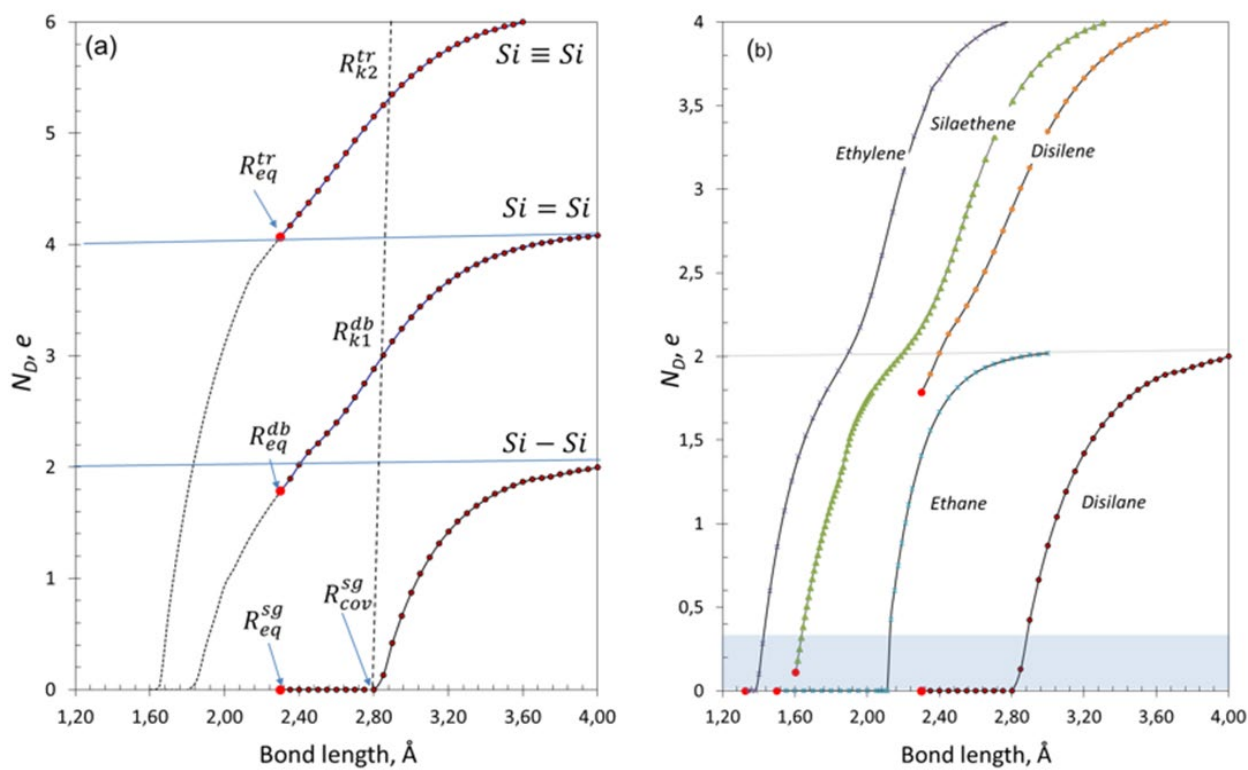

**Figure 4.** $N_D(R)$ graphs of silicon covalent bonds. (a) $sp^3Si - Si$ bond of disilane; $sp^2Si = Si$ bond of disilene; $sp^1Si \equiv Si$ bond of disilyne. Dotted vertical line marks the positions of $R_{cov}^{sg}$. Large red points mark positions $R_{cov}$ in all the cases. (b) $sp^2C = C$, $sp^2Si = Si$, and $sp^2Si = C$ bonds of ethylene, disilene, and silaethene, respectively, and $sp^3C - C$, and $sp^3Si - Si$ bonds of ethane and disilane Light blue band marks dispersion of the $N_D$ values characteristic for $sp^2$ nanocarbons. UHF AM1 calculations.

As can be seen from Figure 4, all multiple covalent bonds involving silicon are radicalized, the $sp^2Si = C$ bond to the least extent and the $sp^1Si \equiv Si$ bond to the greatest extent, and only the single $sp^3Si - Si$ bond of disilane reliably binds neutral atoms. This tendency of interatomic interaction is clearly visible in modern silicon chemistry. Thus, individual disilanes and disilane components of more complex structures constitute the main content of empirical silicon and organosilicon chemistry. Disilenes turned out to be extremely chemically active, as a result of which, from time to time, it is possible to stabilize them only in the form of dimeric inclusions in other molecules [76,77]. In contrast, silaethenes are successfully synthesized and analyzed [78,79]. Empirical evidence for the observation of silicene, a silicon analogue of graphene, is a consequence of the erroneous classification of real $sp^3$ structures of silicon on substrates as $sp^2$-type structures (see a detailed discussion of the issue in [74,75]). Individual disilynes could not be synthesized, and the triple $sp^3 Si \equiv Si$ bond was only detected in the $-Si-Si \equiv Si-Si-$ chain [80, 81].

## 4. Spin Chemistry of Alkane Bonds

As follows from the analysis conducted in terms of bond lengths, alkane bonds generated by classical covalent chemistry remain faithful representatives of this chemistry, regardless of the causes and nature of their origin. Single carbon $sp^3 C - C$ bonds form the basic structural framework of organic compounds. One example of their consolidated operation is the formation of cord-like backbones of organic polymers. Figure 5 demonstrates the action occurred in the course of virtual free-radical polymerization of styrene [82-84].

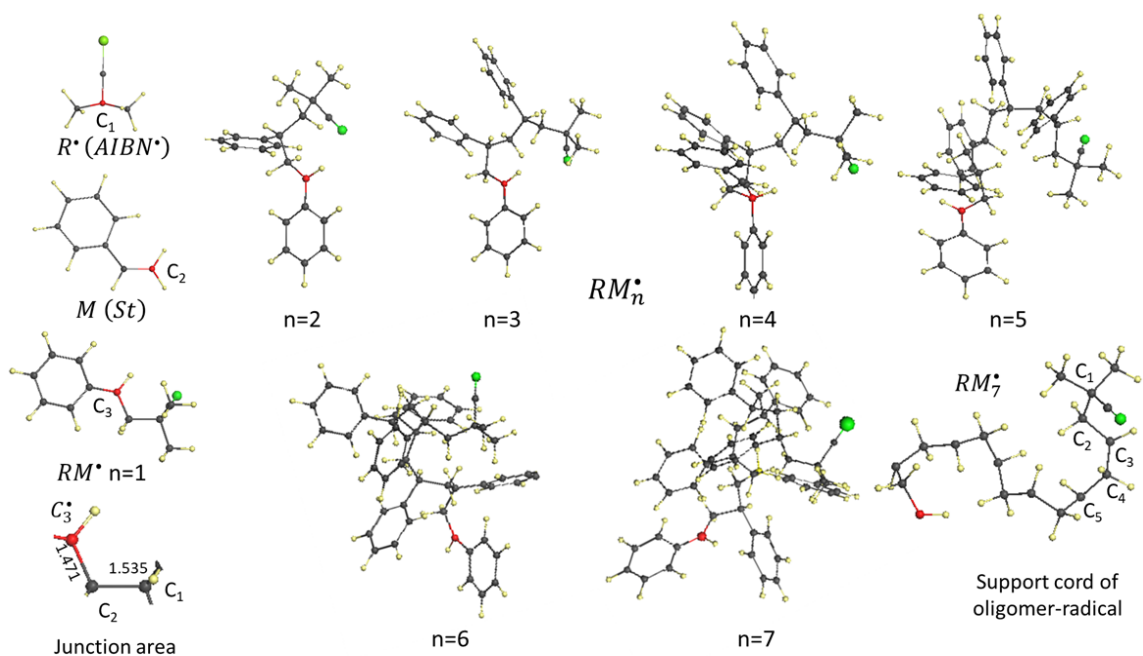

**Figure 5.** Digital twins present free radical oligomerization of styrene. Small yellow and gray balls mark hydrogen and common carbon atoms, respectively. Carbon targets are marked red. Larger green balls depict nitrogen atoms. UHF AM1 calculations.

The reaction occurs in a virtual reaction solution consisting of vinyl monomers $M$ (styrene) and free radicals $R^\bullet$. The alkyl radical $AIBN^\bullet$ plays this role. The reaction begins with the formation of monomer radicals $RM^\bullet$. The target atoms of these primary reactants are determined within the framework of the spin theory of radicals briefly described in the Sections 3.1 and 3.2. In the free radical, the target is the central carbon atom $C_1$, whose odd electron appears as unpaired with a partial electron number of $N_{DA}$= 0.807 e. In styrene, the target is the carbon atom $C_2$ of the CH$_2$ unit of the vinyl group with $N_{DA}$= 0.106 e. The radical and monomer are bound by a covalent bond between their target atoms, which leads to the formation of monomer radical $RM^\bullet$ with an active target atom, which is the carbon atom $C_3$ of the CH unit of the styrene vinyl group, the chemical activity of which is $N_{DA}$ = 0.67 e. Intermolecular junction in the $RM^\bullet$ (see image in lower left corner of the figure) is configured by two covalent bonds $C_1 - C_2$ and $C_2 = C_3$, the first of which is a standard alkane bond, while the second retains the character of an alkene one due to the unpaired electron on the $C_3$ atom. The formation of a dimer radical (and each subsequent member of the oligomer radical chain $RM_n^\bullet$) occurs through the formation of a single covalent bond between the carbon atom $C_3$ of the CH component of the vinyl bond of the $RM^\bullet$ (or $C_{2n+1}$ atom of the preceding $RM_{n-1}^\bullet$ oligomer radical) and the CH$_2$ carbon atom of vinyl group of a new monomer, thus building a chain of alkane bonds that makes up the supporting cord of the oligomer molecule. The configuration of this cord in the oligomer radical $RM_7^\bullet$ is shown in the lower right corner of the figure. The cord consists of alternating CH$_2$ and CH groups, the latter of which is suspended by a benzene ring. Each next-added monomer contributes two atoms of its vinyl group ($C_2$ and $C_3$ in the case of the initial monomer-radical $RM^\bullet$) to this cord, forming an alkane bond $C_1 - C_2$ with the preceding free radical and leaving the alkene bond $C_2 = C_3$ in a strained state. The addition of the next monomer to $RM^\bullet$ with the pilot atom $C_3$ transforms the alkene bond $C_2 = C_3$ into an alkane $C_2 - C_3$ and completes the chain with two new bonds - an alkane bond $C_3 - C_4$ and an alkene bond $C_4 = C_5$. The pilot atom, $C_5$, drives polymerization further, replacing the alkene bonds of the oligomers with alkane bonds, leaving only one alkene bond on the ever-elongating chain, marking the pilot carbon atom of the last monomer added. The movement of the pilot alkene bond along the chain, leaving behind only a chain of alkane bonds, is the essence of vinyl monomer polymerization.

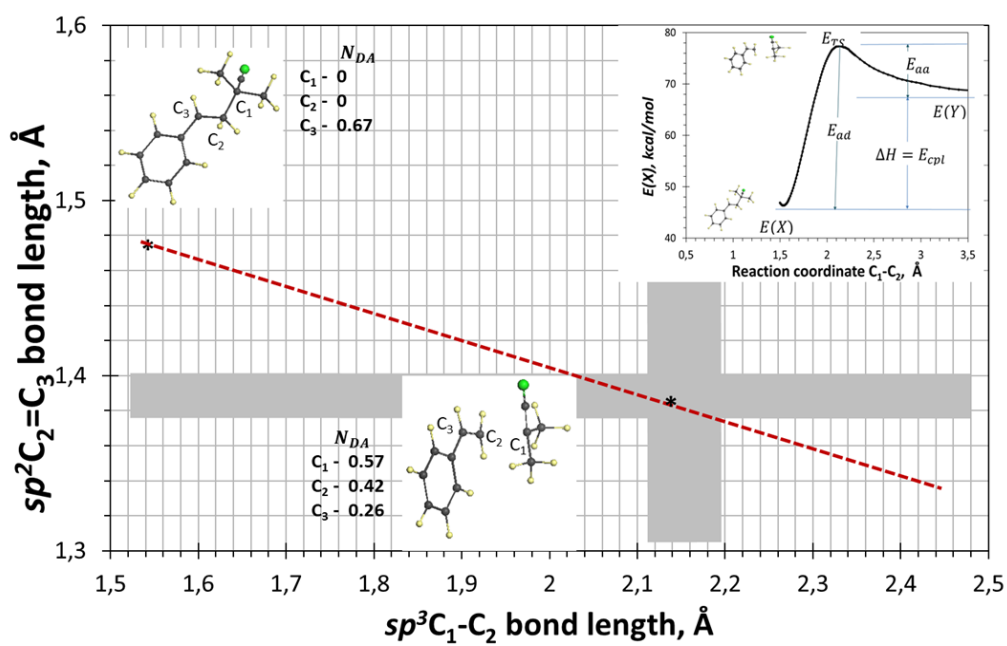

**Figure 6.** The interconnection between the alkene $sp^2 C = C$ and alkane $sp^3 C - C$ bonds, composing the intermolecular junction of the monomer radical $RM^\bullet$. See the junction determination in Figure 5. Inset presents the energy graph $E(R)$ describing the virtual dissociation of the $RM^\bullet$. Light gray bands mark the dispersion of the $R_{cov}^{sg}$ (vertical) and $R_{cov}^{db}$ (horizontal). Images are snap-shots corresponding to structures depicted with black points. UHF AM1 calculations.

As described above. the pilot alkene bond is tightly interconnected with alkane bond that couples new monomer with the current oligomer radical [82]. The feature is presented in Figure 6 for the case of monomer radical $RM^{\bullet}$. Red broken line depicts the elongation of the $sp^3C_1 - C_2$ bond that connects free radical and monomer in the course of the $RM^{\bullet}$ dissociation. The dissociation starts at the (x,y) point (1,535; 1,471) indicating that the junction consists of a standard $sp^3C_1 - C_2$ bond and elongated $sp^2C_2 = C_3$ one. Because of the elongation over $R_{cov}^{db}$, the $sp^2$ bond is radicalized thus providing generation of effectively unpaired electrons of the total number $N_D$ = 1,925 $e$. The latter causes the generation of the junction target atom C$_3$ with $N_{DA}$= 0,67 $e$. The $N_D$ value is in agreement with a general law that governs a gradual radicalization of the $sp^2C = C$ bonds presented in Figure 3a. When the elongation of the $sp^3$ bond proceeds, its length approaches $R_{cov}^{sg}$= 2.11 Å, while the $sp^2$ bond gradually shortens up to $R_{cov}^{db}$. Similar to the dispersion of $sp^3C - C$ bonds lengths in the vicinity of $R_{cov}^{sg}$ (see Figure 3b), $R_{cov}^{db}$ is dispersed responding to the changing the bond surrounding. The dispersion zones for $R_{cov}$ values for both bonds are presented with vertical and horizontal light gray bands, related to the $sp^3$ and $sp^2$ bonds, respectively. The intersection of the bands determined the area of transition state indicated on the energy graph presented in the inset of Figure 6. The graph describes the elementary reaction $RM^{\bullet} \rightarrow M + R^{\bullet}$ discussed above and reveals the transition state occurring at the moment of cleavage of the $sp^3C_1 - C_2$ bond that is the reaction coordinate under consideration. This moment is the transition point for this bond from classical to spin covalence. Molecular compositions of the transition state represent open-shell electron systems, so their quantum-chemical analysis requires the use of methods capable of correctly considering the spin-radical interaction of electrons. Regarding the discussed $sp^3$ bond, the presented analysis allows us to conclude that every alkane covalent bond, classical always and everywhere, becomes spin-covalent at the moment of its destruction (or creation), which is always present in chemical reactions marking transition state location in the space of reaction coordinates.

## 5. Spin Covalence of Alkene Bonds

Alkene bonds are present in a significant number of covalent compounds. These can range from segregated bonds, such as those found in vinyl monomers discussed above, to chains of bonds forming a closed $sp^2$ framework, as in the case of polyaromatic hydrocarbons (PAHs) and $sp^2$ nanocarbons, including fullerenes, carbon nanotubes, and graphene domains. The nature of the bonds determines all the characteristic properties of these substances, and this could be discussed endlessly. We will limit ourselves to those that relate to the unique spin nature of these bonds, and will primarily use the quantities $N_D$ and $N_{DA}$, which determine the degree of bond radicalization, as numerical characteristics.

## 5.1. Aromatic Hydrocarbons

A computational experiment performed for 14 n-PAHs using a number of different CI approaches, such as UMP2, QCISD, as well as UHF and UDFT [85], showed that the species become open-shell molecules starting with naphthalene (n = 2). The spin contamination values $\Delta \hat{S}^2$, indicating the open-shell nature of PAHs, are in good agreement with each other for the first three methods, but remain zero and indicate the closed-shell nature of these molecules in the UDFT case for all n. Being based on wave functions, the UHF approximation is clearly well suited to describing the effects of broken symmetry, while the DFT formalism is less suited to considering the subtle features of the correlation of electrons with different spins, as has been repeatedly noted in the literature [86-88].

Thus, for the already mentioned set of n-PAHs, Figure 7a presents the calculated $N_D$ values obtained using the UHF algorithms and the density matrix renormalized group (DMRG) method [89]. As can be seen from the figure, both data sets are virtually identical. Moreover, the linearly increasing dependence of $N_D(n)$ on the number of benzenoid rings in the molecules perfectly explains the difficulty of experimental synthesis of long n-PAHs, among which pentacene (n = 5) is the last well-characterized polyacene. As it turns out, higher polyacenes are indeed very reactive, as a result of which PAHs with n = 6, 7, and 8 can exist either only in a modified form, when additionally introduced protective groups inhibit the high chemical activity inherent to these molecules, or in crystalline neutral matrices at very low temperatures. Molecules with n>8 are not amenable to chemical synthesis at all. While empirical confirmation of the correctness of UHF LDOS distributions in open-shell molecules still faces significant difficulties, requiring significant refinement of existing experimental methods [90], $N_{DA}$ maps are reliably reproduced by scanning open-shell molecules atom by atom in non-contact atomic force microscopy (AFM) experiments. The greatest success has been achieved with microscopes with an oxygen atom at the tip of the probe needle (see [91-95] and references therein). Figure 7b shows AFM images of two aromatic molecules, pentacene and olympicene, consisting of benzene rings. The experiment was conducted with a gold needle with a CO molecule at its tip [91]. With the color scheme used in the experiment, brightly colored markings indicate atoms of the molecule that interact weakly with the tip. Conversely, the most shadowed atoms interact with it most strongly.

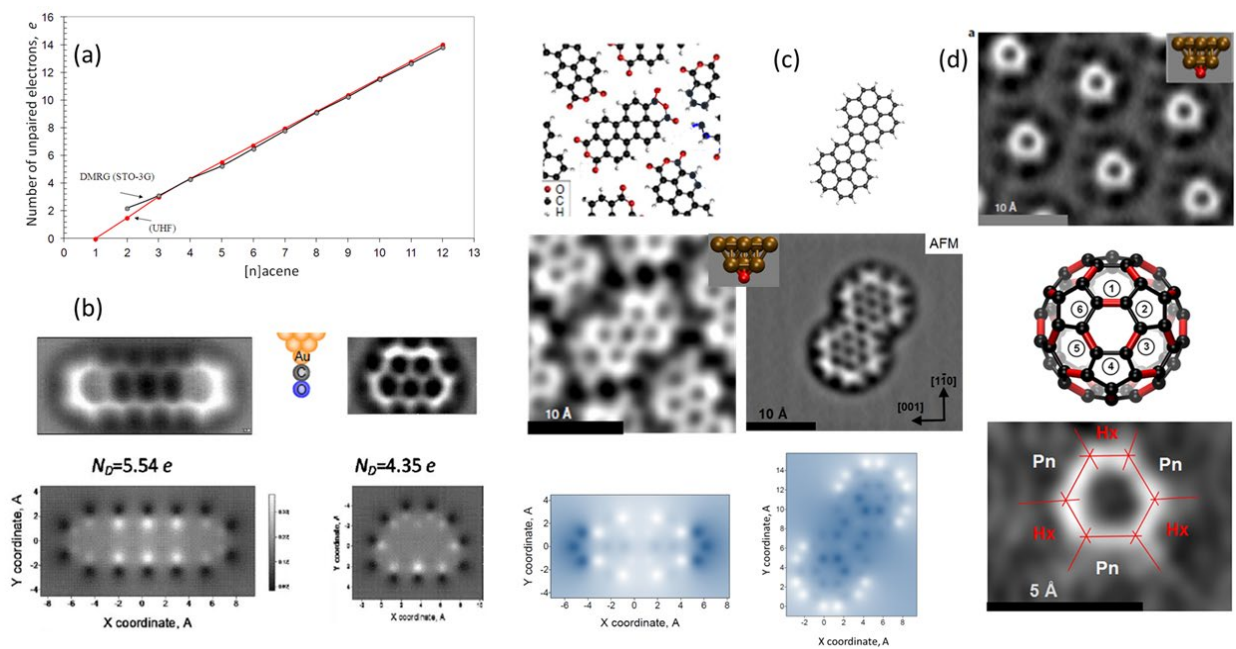

**Figure 7**. (a) Total number of effectively unpaired electrons $N_D$ in polyacenes calculated by using DMRG (STO-3G)) [89] and UHF AM1 formalism. No data scaling. (b) Atomic-resolved AFM image (top) and the ACS $N_{DA}$ maps (bottom) of pentacene (left) and olympicene (right). (c) Digital twin (top), AFM images under the CuO$_x$ tip (middle), and ACS $N_{DA}$ maps (bottom) of 3,4,9,10-perylene-tetracarbonyl-dianhydride (left) and dicoronulene (right) (see text). UHF AM1 calculations. (d) AFM images (top and bottom) and digital twin (middle) of fullerene C$_{60}$ [95].

In conducting these and other AFM experiments with an atomically sharpened tip, the authors aimed to obtain atomic-resolution images of molecules. The fact that these images turned out to be uneven in brightness was unexpected and was presented as experimental evidence of differences in the interactions of the tip's terminal oxygen atom with the atoms of the scanned molecule. The reason for this difference was not discussed until the author of the current article proposed to associate this difference with the ACS $N_{DA}$ maps of the corresponding molecules. The $N_{DA}$ maps of these molecules calculated by the author are presented at the bottom of Figures 7b and 7c [71]. The color scheme of the virtual images in Figure 7b is inverted with respect to the experimental one. A comparison of the empirical images and virtual maps reveals that only carbon atoms are visible experimentally in AFM experiments. As for the calculated $N_{DA}$ values, they are fully confirmed experimentally, correctly ranking the strong and weak interactions of the microscope tip with the molecule under study when scanning their atoms. Thus, AFM with atomic resolution is indeed a reliable method for reading ACS $N_{DA}$ maps.

Figure 7c shows the results using a different microscope tip configuration [95]. In this case, we are dealing with a copper tip delicately oxidized, so that an oxygen atom is located at end of the $CuO_x$ tip. The figure presents the results for 3,4,9,10-perylene-tetracarbonyl-dianhydride adsorbed on the Ag(111) surface under Van der Waals attraction. The right side of the figure presents similar results for the dicoronulene molecule, also adsorbed on the Cu(110) surface. The brightness of carbon atoms in AFM is proportional to the interaction strength of the oxygen atom of the tip with the atoms of the molecule. The comparison of empirical images and virtual ACS $N_{DA}$ maps confidently confirms the earlier conclusion that atomic-resolution AFM is indeed a reliable method for reading ACS $N_{DA}$ maps. Completing the AFM image in Figure 7d is a pinpoint image of $C_{60}$ fullerene molecules adsorbed on the Cu(111) surface and held there by Van der Waals forces [95]. The details of this image will be discussed in the next section.

## 5.2. Fullerenes

An analysis of the structural data for the molecule [96] showed that the diffraction pattern was consistent with the assumption that the proper digital twin of the molecule is a truncated icosahedron with point group $I_h$ symmetry, which is formed by two types of $sp^2 C = C$ bonds, one of which, separating two hexagons, has a length of 1.398 Å (short bonds), while the other, forming pentagons, has a length of 1.455 Å (long bonds). The same pattern was confirmed by a diffraction study of $C_{60}$ crystal, which again revealed two types of bonds differing in length [97]. Based on the general concepts of $C \leftrightarrow C$ bonds at that time, the former were associated with double $sp^2 C = C$ carbon bonds, and the latter with single $sp^3 C - C$ bonds. Within the framework of this concept of the molecular structure, experimental data related to neutron diffraction from $C_{60}$ powder [98], NMR data [99], and repeated X-ray diffraction data for a $C_{60}$ crystal [100] were further explained. Numerous studies of STM images of $C_{60}$ fullerene adsorbed on semiconductor surfaces [101,102] helped confirm the icosahedral (nearly spherical) shape of the molecule. The molecule is most fully represented using high-resolution non-contact AFM with an oxygen atom at the tip of the probe (CuO needle) [95] in Figure 7d. As can be seen from the figure, the molecule is indeed a truncated icosahedron with two types of interatomic bonds.

Numerous calculations of the molecule' digital twin confirmed its icosahedral shape, characterized by two sets of bonds, but disagreed on the definition of symmetry. According to RHF and DFT calculations, the molecule's symmetry is $I_h$, while UHF shows $C_i$. Recall that in the UHF formalism, the reduction in symmetry concerns spin symmetry, but it affects subtle characteristics of the geometric shape as well. In the case of the $C_{60}$ fullerene, the latter concerns the length dispersion of the $sp^2 C = C$ bonds that form the molecule's framework. Figure 8a (top) shows the bonds distribution for two structures of $C_{60}$ digital twins that correspond to the RHF and UHF solutions, while Figure 8a (bottom) details the distribution of these bond lengths in each of the structures [66,67,71]. As can be seen from the figure, while the average bond lengths are virtually identical for long bonds and differ slightly for short bonds, the corresponding dispersions caused by the splitting of both types of bonds into four groups differ by four and 16 times for long and short bonds, respectively. This large difference in dispersion is an adequate response of the structure to a change in the spin state. As can be seen from Figure 8a, all long $sp^2 C = C$ bonds of fullerene $C_{60}$, and some of the short ones, are located above the critical value $R_{cov}^{db}$ = 1.395 Å, so the significant number of effectively unpaired electrons $N_D$, amounting to 9.84 e, seems quite natural.

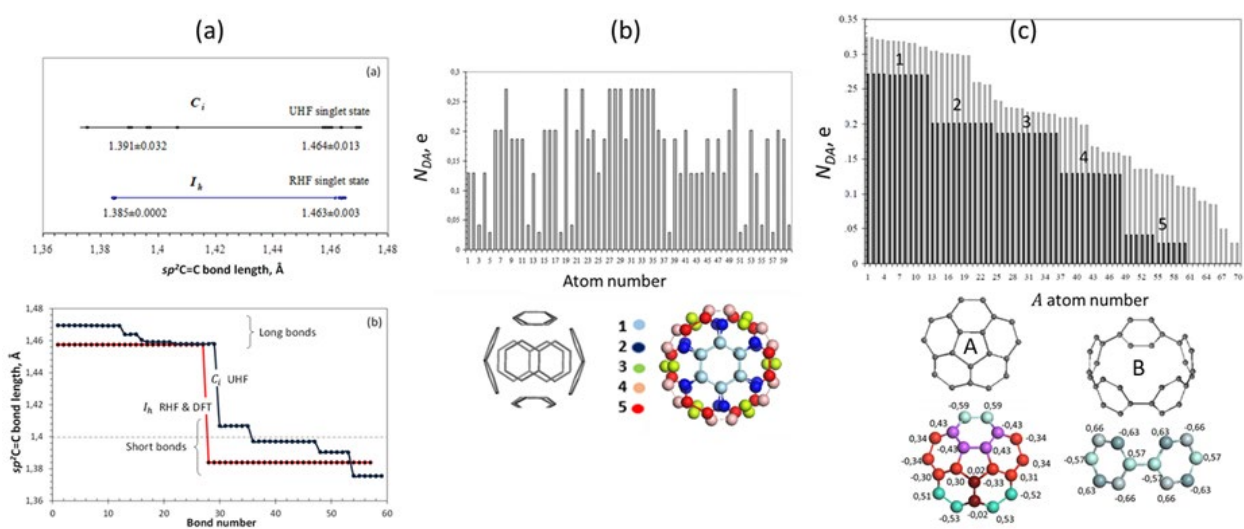

**Figure 8**. $sp^2$ Fullerenes in light of the spin-radical concept. (a) Space structure in terms of covalent bonds. (b) and (c) Effectively unpaired electrons of $C_{60}$ and $C_{70}$, respectively. UHF AM1 calculations.

The fragmentation of short and long bonds into groups is expectedly reflected in their spin characteristics. As can be seen from Figure 8b (top), the ACS $N_{DA}$ graph over atoms of the molecule $C_{60}$ also has a grouped appearance, clearly evident in the $Z \to A$ ordering of the values (see the black histogram in Figure 8c (top)). A similar pattern is observed for effectively unpaired electrons of fullerene $C_{70}$ with a total number $N_D$= 14.4 e. In the case of $C_{60}$ and $C_{70}$ molecules, both $N_D$ values constitute about 15-20% of the total number of odd electrons. As for $Si_{60}$ with $N_D$= 63.52 e, not only all 60 odd π electrons but also some σ electrons are unpaired, which evidences about a complete radicalization of the molecule.

In terms of the spin-radical concept, the discussed ACS $N_{DA}$ graphs in Figures 8b and 8c exhibit chemical activity of the molecules' atoms. The $Z \rightarrow A$ formatted graph of fullerene $C_{60}$ manifests five groups of 12 atoms each, which are characterized by the same $N_{DA}$ value within the group. Each group atoms forms six identical $C_2$ pairs held by one of the $sp^2 C = C$ bonds. The spin densities of the atoms in any pair are equal in magnitude and opposite in sign, so that the spin density of the molecule is zero [103]. According to the ACS $N_{DA}$ graph, the $C_{60}$ molecule consists of six identical $C_{10}$ compounds formed by five pairs corresponding to the total number of groups. Distributing the atoms across six fragments according to the map shown in the figure, we obtain a 6*$C_{10}$ configuration consisting of six identical naphthalene nuclei. Applying different colors to atoms with different $N_{DA}$ values, we obtain a multi-colored 'chemical portrait' of the $C_{60}$ molecule in the singlet state in the UHF approximation [66-68, 71, 104] shown in Figure 8b (bottom). The same numbering concerns five groups indicated in Figure 8c (top) and five colors in Figure 8b (bottom). Accordingly, two hexagons formed by light blue atoms manifest the most active area of the molecule. Any first addition reaction with $C_{60}$ starts with one of these atoms.

$Z \rightarrow A$ formatted ACS $N_{DA}$ graph of the molecule $C_{70}$ shown in Figure 8c (top) is much less contrasted in comparison with that of $C_{60}$. Nevertheless, as previously, the graph points to a well-defined grouping. The $D_{5h}$ symmetry of the molecule in the UHF singlet state governs the molecule structure decomposition into three five-benzenoid fragments. Two 20-atom fragments A, one of which is shown in in the middle of the figure, are formed by conjugated benzenoids and look like a five-lobe flower each. Five benzenoids mutually coupled via a single $sp^2$ bond form a 30-atom closed rarefied chain-bracelet B. The highest $N_{DA}$ values are concentrated just in this area. It is clearly seen in Figure 8c (bottom), where a color chemical portrait of $C_{70}$ is presented, but in terms of spin density that is fully consistent with $N_{DA}$ values [68,69]. Any first addition reaction on $C_{70}$ occurs on the chain-bracelet atom.

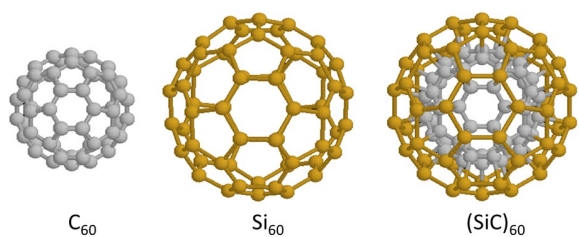

$C_{60}$      $Si_{60}$      $(SiC)_{60}$

**Figure 9.** Digital twins of carbon-silicon fullerenes. UHF AM1 calculations.

Figure 9 returns us to the comparison of carbon and silicon fullerenes. The theory of spin-radical interactions allows only the $C_{60}$ fullerene to exist, thereby defining, in agreement with empirical data, the boundaries of the region of permissible $N_D$ per-one-bond values. This region is marked by the light-gray band in Figure 4b. The relatively small $N_D$ values determine the status of fullerene $C_{60}$ as a stable radical. The $N_D$ corresponding to $R_{eq}^{db}$ for the $sp^2 Si = Si$ bond lies well above this region, once again confirming the impossibility of the $Si_{60}$ fullerene existence. However, as Figure 4b shows, the $N_D$ value corresponding to $R_{eq}^{db}$ of the $sp^2 Si = C$ bond falls within the region of permissible $N_D$ values. Thus, the spin-covalent chemistry of the $sp^2 Si = C$ bonds does not hinder formation of the $(SiC)_{60}$ fullerene, which may allow it to be synthesized in the future.

## 5.3. Carbon Nanotubes

From the vast ocean of experimental and virtual data on carbon nanotubes (CNTs), we select only those related to the molecular structure of CNTs, which is based on chains of $C \leftrightarrow C$ covalent bonds. Although diverse in their physical and chemical properties, CNTs are uniform from a molecular perspective. They all represent cylinders of varying diameters and lengths, formed by twisting a monatomic honeycomb web. The benzoid rings of the web are formed by $sp^2$ bonds with lengths ranging from 1.39 to 1.45 Å, and their dispersion can vary. Most bonds are longer than the critical length $R_{cov}^{db}$, resulting in the electron system of the tubes containing a significant $N_D$ number of effectively unpaired electrons. Available virtual data show that the $N_D$ per-one-bond values fall within the region of permissible data, indicated in Figure 4b, evidencing a stable-radical character of the species.

The UHF approach was first applied to two fragments of (4,4) defect free and (4,4) 5–7 defect single-walled CNTs (SWCNTs) [105]. The obtained characteristic ACS $N_{DA}$ graphs along the tube's atoms, manyfold supported in further studies [106], are shown in Figure 10. The graphs reliably monitor the tube atoms' atomic chemical susceptibility as well as tight connection of the latter with the tube structure, thus highlighting the distribution of the $sp^2 C = C$ bond length excess over $R_{cov}^{db}$ along the tube. The discussed graphs are related to CNT digital twins in the form of fragments of (4,4) SWCNT, equilibrium structures of whose digital twins are shown in the figure. One end of all fragments is capped, while the other is either open but hydrogen terminated (NT1) and empty (NT2), or capped (NT3). The total number of effectively unpaired electrons $N_D$ of the tubes is concentrated around 32 $e$. The distribution of these electrons over the tubes' atoms forms characteristic ACS $N_{DA}$ graphs that are shown in in the figure. As seen in Figure 10a, the tube can be divided into three regions. The first is related to the cap with adjacent atoms. The second concerns mainly the tube sidewall while the third refers to the open end terminated by hydrogen atoms. The biggest nonuniformity of the $N_{DA}$ distribution is characteristic of the cap region. One should pay attention to the fact that the largest $N_{DA}$ values belong to atoms that form the longest bonds. As for the sidewall region, the $N_{DA}$ distribution is practically uniform, with the $N_{DA}$ value scatter not bigger than 0.5%, which is consistent with a quite uniform distribution of the bond lengths as well. The $N_{DA}$ graph in the end region is significantly affected by the hydrogenation. The curve with dots in the figure presents the free valence of the tube atoms that well coincides with the ACS expressed by $N_{DA}$. According to the latter, the tube cap is the most reactive part, while the tube sidewall is more passive, with ill-pronounced selectivity along the tube.

Removing hydrogen atoms at the tube's open end, one obtains the $N_{DA}$ graph of the NT2 fragment shown in Figure 10b. A tremendous contribution of the end atoms obviously dominates the graph. This is due to the fact that the $sp^2 C = C$ bonds are replaced in the region with the $sp^1 C \equiv C$ ones. The transformation naturally results in increasing the total number of effectively unpaired electrons $N_D$ from 32.38 $e$ to 39.59 $e$. The injection of additional effectively unpaired electrons disturbs the $N_{DA}$ graph of the hydrogen-terminated tube (shown in the figure by bars) quite considerably. It is important to note that the changes occur not only in the vicinity of the open end, but also in the opposite cap end. Practically no changes occur along the tube sidewall, which seems to serve as a peculiar resonator for the electron conjugation. Addressing the chemical activity of the tube, dominant activity of the empty-end atoms is evident. Placing the cap at the open end of the tube (NT3 fragment) makes its ACS $N_{DA}$ graph almost symmetrical (see Figure 10c).

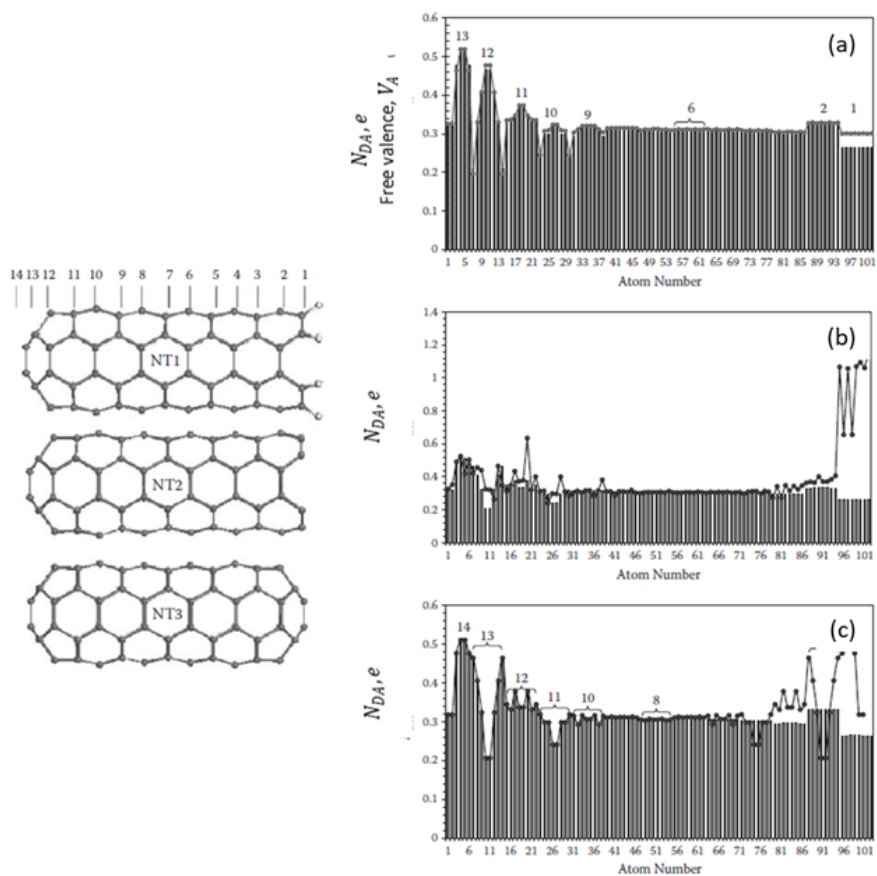

**Figure 10.** Digital twins of fragments of the (4,4) SWCNT and their ACS $N_{DA}$ graphs. (a), (b), and (c) are related to NT1, NT2, and NT3 species, respectively. UHF AM1 calculations

Exhibited peculiarities of the discussed ACS $N_{DA}$ graphs are generally common for all types of CNT and allow making general conclusions concerning addition reactions to be expected [106]. The space of chemical reactivity of any SWCNT coincides with its coordinate space, while remaining different for particular structure elements. Increasing the tube length leads to a broadening of zone 2 in its ACS $N_{DA}$ graph without changing its amplitude. Increasing the tube diameter affects the graph' amplitudes if it is accompanied by a change in the dispersion of bond lengths within the tube. The apex and end atoms of the tubes are sensitive to the heteroatomic state of the surrounding environment. Local additions of short-length addends (involving individual atoms, simple radical, and so forth) to any SWCNT are the most favorable at open empty ends. Following these places in activity are end caps, defects in the tube sidewall, and the sidewall itself.

**5.4. Graphene Domains**

Graphene is currently one of the most actively studied objects of modern material science. Countless theoretical and experimental studies have already been performed, targeting electronic, magnetic, thermal, optical, structural, and vibrational properties. Low and homogeneous chemical reactivity of atoms throughout a graphene sheet is usually expected by the predominant majority of scientists dealing with graphene chemistry. However, the first UHF calculations [107] showed that this is not the case, since the equilibrium length $R_{eq}^{db}$ of about half of $sp^2 C = C$ bonds of graphene domains exceed the critical value $R_{cov}^{db} \cong 1.395$ Å. The domains are significantly radicalized, which is manifested with significant values of both $N_D$ and $N_{DA}$. Research-friendly rectangular digital twins of nanosize graphene domains (nanographenes) are nominated as ($n_a$,$n_z$) NGrs following [108]. Here $n_a$ and $n_z$ match the number of benzenoid units along the armchair and zigzag edges of the domains, respectively. Similar to CNTs, ACS $N_{DA}$ graphs of graphene domains are quite typical, only slightly dependent on the domains size, while $N_D$ grows with the size.

Characteristic ACS $N_{DA}$ graphs for (5,5) NGrs with bare and hydrogen-terminated edges, presented in Figure 11, demonstrates a rather significant variation of the quantity over atoms. As seen in the figure, the highest $N_{DA}$ are characteristic of carbon atoms at the zigzag edges, while those of the armchair edges are similar to the $N_{DA}$ values of atoms of the domain basal plane. The latter are comparable with those of fullerenes (ca. Figure 8) and SWCNT sidewalls (Figure 10). When hydrogen terminators are removed, $N_{DA}$ values on both zigzag and armchair edges grow significantly, still conserving bigger values for zigzag ones. The obtained results alongside with many other known today [69,109,110] made allowance for the following conclusions concerning chemical reactivity of graphene materials. All reactions with graphene domain that is a spacious object are of topochemical nature. There is still no reliable algorithm for such reactions, although the main control element is unlikely to deviate in appearance from the well-tested form of an ACS $N_{DA}$ graph. For now, we have to limit ourselves to this graph of ordinary chemistry when planning the course of successive reaction steps, as in the case of fullerenes and CNTs. Based on the graphs presented in Figure 11, it can be stated that any chemical addend will first be attached to the graphene zigzag edges, both hydrogen- and other heteroatom-terminated and empty. Chemical reactivity of basal-plane atoms depends on the edge termination and cannot be strictly correlated with activity of fullerenes and SWCNT sidewalls. The steadiness of stable radical status for the graphene domains is ensured by the state of edge atoms of the domain.

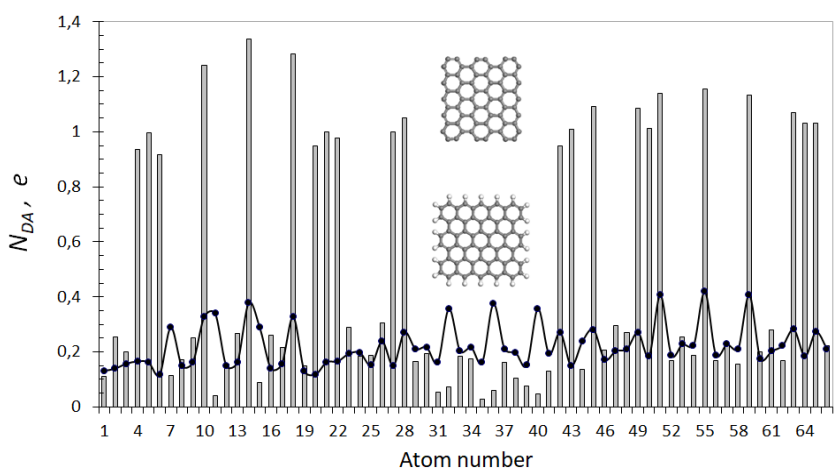

**Figure 11.** Digital twins of bare and hydrogen-terminated (5,5) NGrs and their ACS $N_{DA}$ graphs. UHF AM1 calculations

## 6. Spin Covalence of Alkyne Bonds

Comparing the ACS $N_{DA}(R)$ graphs of ethylene and propyine shown in Figure 3a, one can easily conclude that molecules with alkyne bonds should be more active. In practice, the situation is more complicated. Not many substances with alkyne bond chains are known and graphpolyines (GPYs), as related to modern graphene science, are of a particular interest. The species were object of numerous studies during last time (see comprehensive reviews [21,111-113] that discuss and summarize the state-of-the-art research of the issue. Introduced in Section 1.2, these allotropes are flat one-atom-thick carbon networks, which can be constructed by replacing some $=C=C=$ bonds in graphene by uniformly distributed polyacetylenic linkages $-C\equiv C-C\equiv C-C\equiv C-$ ... ($n$–[–C≡C–]) forming GPYs, among which there are graphynes (GY, $n$=1), graphdiynes (GDY, $n$ = 2), graphtriynes (GTY, $n$ = 3), and so forth. GPYs are largely variable structures, a particular part of which is occupied by species consisting of networks that include $C_6$ hexagons interconnected with polyacetylenic linkages. The main impression on promising interesting properties of GPYs as well as on their possible applications was provided virtually, while experimental evidences are rather scarce. However, almost all the computations were performed in closed-shell approximations without taking into account spin covalence of alkyne bonds. To compensate this drawback let look at some basic components of GPYs from the viewpoint of UHF formalism.

As seen in Figure 12a, diphenyl-$n$-acetylenes ($n$-DPHAs, phenyl terminated polyynes) consisting of two phenyls connected with a varying number of acetylenic linkages (ligaments below) are the basic GPY components. A set of $n$-DPHAs with $n$ from 1 to 4 is shown on the left panels of the figure, while right panels display the relevant ACS $N_{DA}$ maps that exhibit the presence of chemical reactivity of the molecules and disclose its distribution over the molecules' atoms. Going from the top to bottom, one can see how the reactivity map changes in value and space when the acetylene linkage increases. The linkage of one acetylenic unit between benzene rings causes a considerable elongation of some of the ring bonds, thus promoting a significant radicalization of the rings as presented at the right-hand panel. The total number of effectively unpaired electrons $N_D$ = 1.22 $e$ with fractional $N_{DA}$ values on the ring carbon atoms from 0.10 $e$ to 0.08 $e$. Similar small radicalization of $N_{DA}$ = 0.06 $e$ concerns the acetylenic unit. Inclusion of one more acetylenic unit between benzene rings promotes a remarkable elongation of both triple bonds, which, in turn, results in the enhancement of their radicalization just lifting both $N_D$ and $N_{DA}$ values while characteristics of benzene rings remain practically unchanged. This trend is preserved with further increase in the number of triple bonds. As seen in the figure, in due course of growth of the triple bond number, the chemical reactivity of DPHAs is increasingly concentrated on the atoms of acetylenic units evidencing the growing elongation of the latter. This might be explained by the transformation of a quite rigid sole triple bond to a flexible chain of the bonds, which readily promotes the bond elongation, with quite regular spacing of 1.217–1.225 Å within the acetylenic units and of 1.327–1.332 Å between them.

Presented in Figure 12a exhibits a great increase in the chemical activity of polyynes when acetylenic chains becomes longer. Apparently, the finding explains the failure in the synthesis of linear ynes with n >11 [114]. Experimentally observed structures of polyynes with $n$ from 2 to 11 manifested regularly distributed acetylenic units over the chain with characteristic interatomic spacing of 1.201–1.211 Å and 1.353–1.364 Å depending on the length and termination of the acetylenic chains. The data are in perfect agreement with those related to the DPHAs shown in the figure. It proved possible to overcome high chemical activity of long polyynes only placing them in a confining reactor [115]. The linear carbon chains were encapsulated in and protected by thin doublewalled carbon nanotubes. Exceptionally long and stable chains composed of more than 6,000 carbon atoms were obtained.

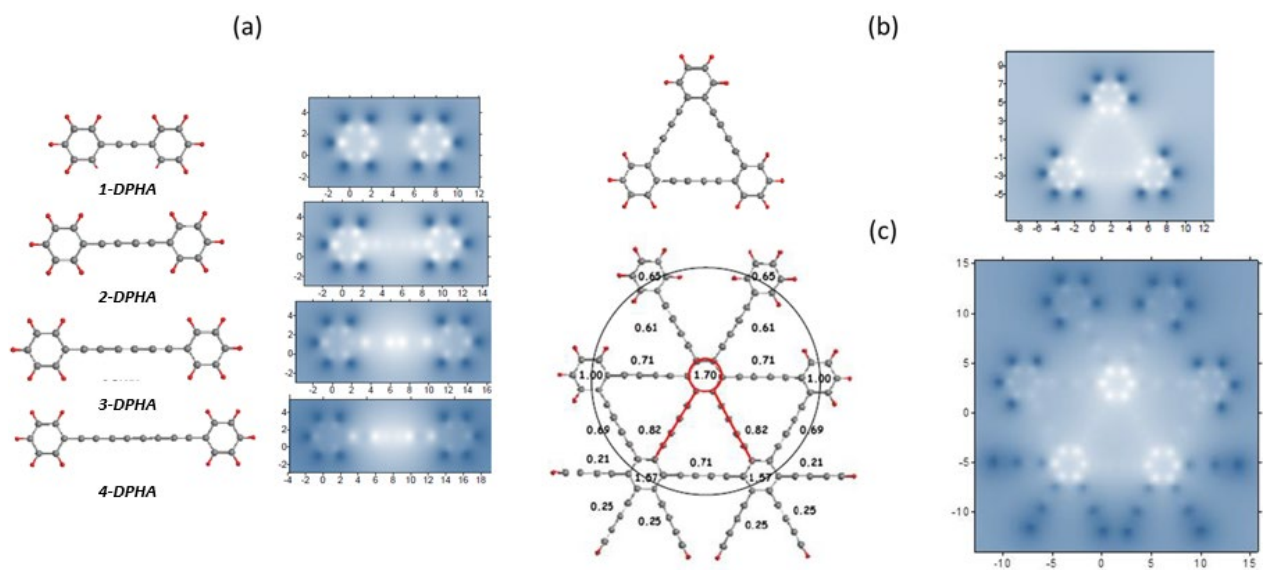

**Figure 12.** Equilibrium structures (left) and ACS $N_{DA}$ image maps (right) of graphpoyynes digital twins. (a) *n*-diphenylacetylens; *n*=1, 2, 3, 4. The maximum value of the map intensity scale varies from 0.10 e for *1*-DPHA and *2*-DPHA to 0.16 e for *3*-DPHA and 0.30 e for *4*-DPHA. (b) One-triangle *2*-DPHA-based compositions with two-branched hexagons $C_6$. (c) 'Irish lace' GDY pattern. The maximum value of the map intensity scale is 0.23 e in (b) and 0.30 e in (c). Figures present summarized $N_{DA}$ values related to hexagons $C_6$ and ligaments. Gray and red balls mark carbon and hydrogen atoms, respectively. UHF AM1 calculations.

The considered DPHAs lay the foundation of various GPYs differing by the number of acetylene linkages between benzenoid rings. Thus, 1-DPHA, 2-DPHA, and 3-DPHA form the grounds of graphyne, graphdiyne, and graphtriyne, respectively. Independently of a concrete structure of the involved DPHA, the composition like a six-petaled flower lays the foundation of the structure of any GPY. Six-branched benzenoid hexagon $C_6$ determines each of the flower centers, while another six hexagons terminate the flower petals. Each of these rings, one-branched previously, gradually becomes six-branched in due course of the GPY growth in plane, which results in a particular triangle pattering of the GPY body consisting of triangle-closed cycles. Three hexagons form the vertices of the triangle while ligaments lie along its sides. Basing on these structural grounds, it is easy to imagine a consequent formation of a regular extended structure of GDY. Thus, Figure 12 b presents one-triangle 2-DPH based compositions. The ACS $N_{DA}$ map, or chemical portrait of the molecule at right-hand panel in the figure impressively exhibits changing in the 2-DPHA atomic reactivity caused by the triangle composition, which reflects a different changing in bond lengths.

A successive ranching of benzenoid hexagons of GDY recalls braiding Irish lace, whose main motive is selected by circle in Figure 12 c. The triangle fragment marked by red constitutes its main part related to the extended lace body. The knitting is obviously a multi-stage complex process, which is difficult to trace in all details. However, taking ACS $N_{DA}$ maps as assistants, it is possible to disclose general trend and regularities for making the following conclusions. GDY presents a large cloth with a regular flower-like print where six-branched benzenoid hexagons play the role of the main floral motif while alkynic ligaments present thin twigs. The motive is a radical, but the status of its radicalization depends on the surrounding structure. The highest radicalization is related to that one surrounded by identical six-branched benzenoid hexagons. The motive main radicalization is concentrated on the hexagon, while each ligament is about half less reactive.

The GDY cloth as a whole is highly radicalized and, consequently, chemically reactive. The $N_{DA}$ values of the motive atoms are similar to those that are characteristic for carbon atoms of fullerenes, nanotubes, and basal plane of graphene domains. Similar to the latter bodies, the $N_D$ per-one-alkyne bond fall inside the light gray band on Figure 12c, because of which GDY are stable radicals and can exist at ambient conditions, once inclined to a variety of chemical transformations. Cutting and saturation with defects will considerably enhance the body reactivity, which should be taken into account when discussing a possible controlling of electronic properties of GDY devices [113, 116].

## 7. Carbon Catalysts in Light of Their Spin Covalence

The catalytic properties of carbon molecules are of interest to two rapidly developing branches of metal-free catalysis: organocatalysis [117-120] and carbocatalysis [121-131]. It should be noted that no bond-classified class of carbon molecules has attracted particular attention in organocatalysis: neither alkanes and alkenes, nor alkynes. In the literature, only a few references can be found to the involvement of alkenes [120] or alkynes [132]. Carbenes [133] have received somewhat more attention. Analysis of numerous manifestations of organocatalysis has revealed a clear correspondence between the catalytic eficiency of molecular catalysts and their acidic and alkaline properties. The release of catalysts from the metals has led to the only mechanism of catalysis associated with the presence of oxygen, sulfur, nitrogen, etc. heteroatoms.

In contrast to organocatalysis, carbocatalysis utilizes purely carbon catalysts, which mainly are representatives of *sp²*nanocarbons, including fullerenes, carbon nanotubes (CNTs), and graphene materials. The latter play a leading role in these reactions, while the use of fullerenes and CNTs is still scarce. Graphene carbocatalysis is divided into two large groups determined by the size of the catalysts used, namely, large-area and small-area ones. Catalysis in the first group is solid-state one, the features and control of which are determined by the surface properties of the reactants participating in the reaction (see [126,134-138] and references therein). The second group, combined with fullerenes and CNTs, is molecular carbocatalysis, a brief analysis of which, in light of the spin properties of their $sp^2 C = C$ bonds, will be presented below. Interested primarily in the virtualization of this process, we begin our discussion by determining the structural dimensions of the catalysts and establishing the nature of their active sites. This primarily concerns graphene materials.

### 7.1. Graphene Carbocatalysis

The main types of molecular graphene structures are shown in Figure 13. Since *sp²* graphene is chemically an atomic system with unsaturated valence bonds, the type of molecular structure is determined by the chemical state of two structurally sensitive regions of the graphene domain— its edge atoms with dangling bonds and basal plane atoms with unsaturated valences. The bare graphene domain is extremely chemically active and does not actually exist in practice. At ambient conditions, valence saturation of the domains occurs in two stages [109]. The first concerns the domain edge atoms, while maintaining the *sp²* configuration of the carbon atoms in the basal plane. Such necklaced molecules are stable radicals [139] and are widely known as the basic building units (BSUs) of all types of $sp^2$ amorphous carbon ($aC$) [37]. Nanoscale synthetic reduced graphene oxide ($rGO$) also belongs to this group. The second stage of valence saturation affects the carbon atoms in the basal plane of the graphene domain, the $sp^2$ configuration of which is replaced by the $sp^3$ one [140]. At 100% valence saturation, the chemical activity of the molecule becomes zero. Graphene oxide ($GO$) belongs to this type of molecule. $GO$ does not exist in nature. It is synthesized by the harsh oxidation of highly fragmented graphite, so that the final product is a nanoscale powder, which is then used to obtain $rGO$ by chemical reduction.

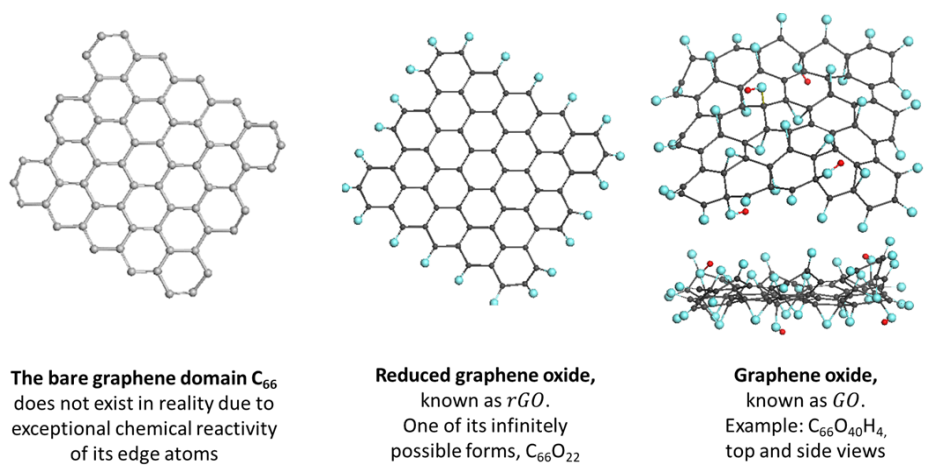

The bare graphene domain $C_{66}$ does not exist in reality due to exceptional chemical reactivity of its edge atoms

Reduced graphene oxide, known as $rGO$. One of its infinitely possible forms, $C_{66}O_{22}$

Graphene oxide, known as $GO$. Example: $C_{66}O_{40}H_4$, top and side views

**Figure 13**. Digital twins of typical graphene catalysts.

The above-mentioned acid-base concept of the active cite of organocatalysts, proposed more than 30 years ago, were transferred to carbocatalysts (see details in review [141,142]). In the latter case, the concept has been somewhat corrected emphasizing a particular role of organic molecules due to many of them, possessing acidity or basicity as well as redox activity, can catalyze reactions. The view on virtual carbocatalysis from the standpoint of the spin chemistry of graphene molecules was firstly proposed in [143] for $aC$ catalysts, BSUs of which were presented with a conglomerate of graphene-oxynitrothiohydride stable radicals. The chemical activity of the BSUs atoms is reliably determined computationally, which allows mapping the distribution of active sites in these molecular catalysts. The presented maps reliably show the BSUs radicalization provided with carbon atoms only, the non-terminated edge part of which presents a set of active cites. Spin mapping of carbocatalysts active cites was suggested as the first step towards the spin carbocatalysis of the species.

Figure 14 presents a comparative view on the radical behavior of the studied $aC$s exemplified by 'chemical portraits' of the relevant digital twins. Each of the exhibited molecules heads a set of digital twins suggested for the studied amorphics [37]. The ACS $N_{DA}$ map of the parent (5, 5) NGr domain accompanied with the $Z \to A$ formatted ACS $N_{DA}$ graph is shown in Figure 14a. Bright spots on the map indicate cites of the most chemical activity. Quantitative presentation of this activity is shown by the ACS $N_{DA}$ graph nearby. As seen in the figure, large $N_{DA}$ values of 0.9÷1.3 e distinguish 22 edge atoms while values $\leq$ 0.3 e belong to basal-plane atoms.

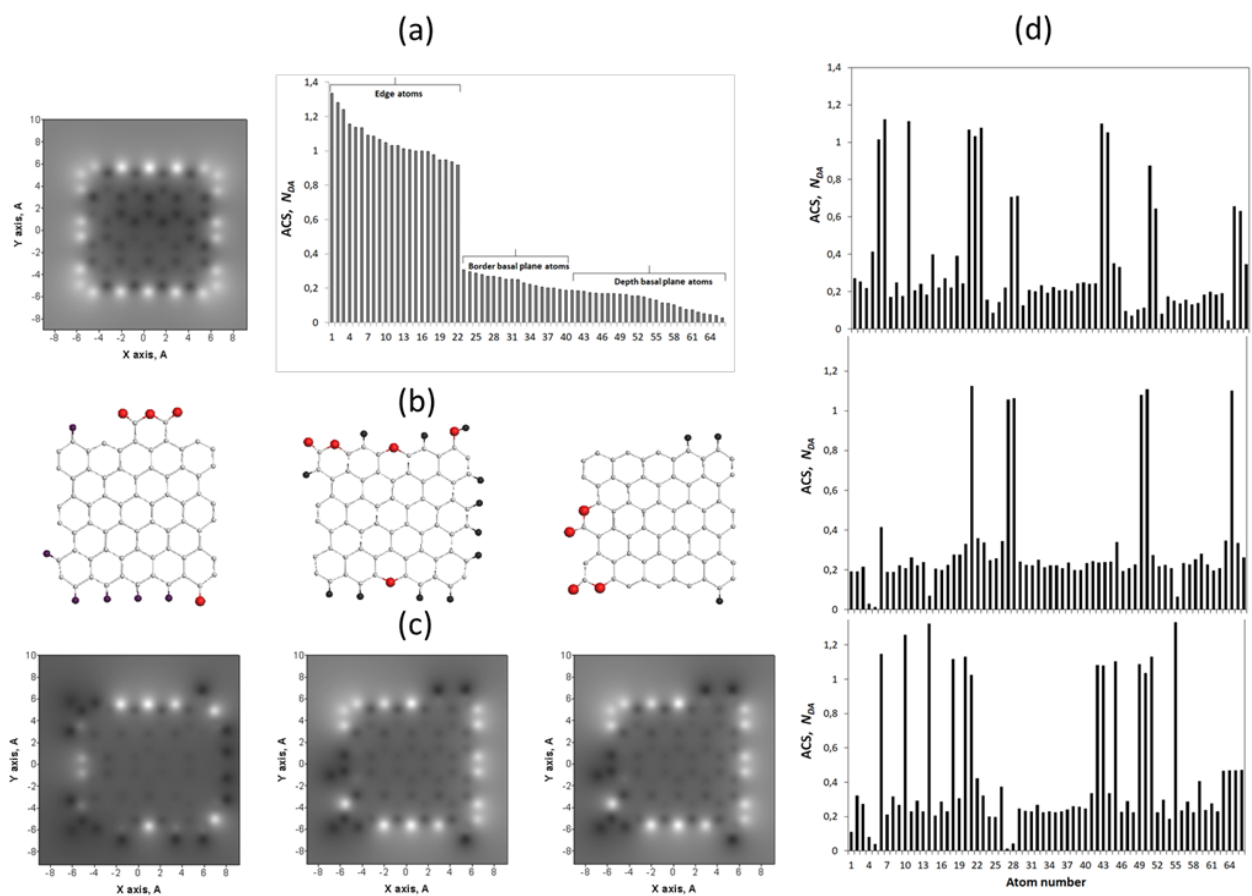

**Figure 14.** Spin character of digital twins of graphene catalysts. (a) Parent graphene domain $C_{66}$. Graphene oxyhydrides $C_{66}O_4H_6$, $C_{66}O_5H_{12}$, and $C_{66}O_4H_3$ from right to left on (b) and (c) and from top to bottom on (d), respectively.

This characteristic feature of bare graphene molecule is generally preserved for all necklaced graphene molecules as well while is evidently disturbed. As visible in Figures 14b, 14c, and 14d, the inclusion of hydrogen and oxygen into the parent domain circumference expectedly inhibits the activity of the edge carbon atoms, directly involved in the new bonding. At the same time, the figures show enhancement of the activity of basal-plane atoms. The feature clearly demonstrates a peculiar collective character of chemical events occurring with graphene molecules [70]. The molecular chemical susceptibility $N_D$ of the parent domain $C_{66}$ (33.49 e) considerably reduces in all the necklaced molecules and constitutes 23.65 e ($C_{66}O_4H_6$), 20.49 e ($C_{66}O_5H_{12}$), and 29.66 e ($C_{66}O_4H_3$). As is visible in the figures, each decoration act causes remarkable changes in the ACS $N_{DA}$ graphs thus revealing the bonds' redistribution of the whole molecules. Despite the effect in each case is expectedly individual, the general pattern of the ACS $N_{DA}$ graph is conserved. As for functional groups involving heteroatoms, a dominant role of which in carbocatalysis has been largely discussed until now, as seen in the figure, they completely lose their activity after attaching the carbon core and are not capable to further lead any chemical reaction while adjacent 'empty' carbon atoms are ready to play the role.

According to one of the most reputable experts in the field, "catalytic applications of carbon materials are as old as the discipline of physical chemistry, and probably even older" [144]. In fact, $aC$'s materials such as activated carbons and carbon blacks have been used for ages in heterogeneous catalysis as either catalysts or catalyst supports. The first documentation of the issue was done about hundred years ago in a report about the aerobic oxidation of oxalic acid occurred on the surface of charcoal [145]. The latest achievements of this biocarbon catalyst can be found in reviews [121,146]. By the end of the nineties, research in the area of $aC$ catalysis was a well-defined field of material science, whose further development is reflected in a large number of publications, thoroughly reviewed in a set of reviews and monographs [122-124,126-130,147,148].

In addition to serving as catalysts, small-area necklaced graphene BSUs of $sp^2$ $aC$s provide a protective function similar to that of the large-area ones [137,138]. This is a reason of the carbon coating that accompanies great number of natural minerals. In the case of the dominant presence of carbon, say, of natural deposits of shungite, the carbon coating is transferred into encapsulating nanosize carbon sacks, whose skin is constructed from shungite's BSUs while the interior is filled with ideally crystalline quartz [149-151]. Averaged size of the quartz particle constitutes (80± 2) nm, which is well correlated with the size of a sphere, constructed with densely packed BSUs of shungite of $\approx$ (2.5 x 2) nm$^2$ in size [37]. As occurred, $sp^2$-carbon sacks were proposed for and realized in biotechnology as well [152], allowing to suspect that the presence of the nanosize graphene's traces in the COVID vaccines [153,154] may be explained by the necessity to elongate the mRNA time of life encapsulating the latter in the carbon sacks.

### 7.2. Fullerene Carbocatalysis

The influence of the presence of fullerene, primarily of fullerene C$_{60}$, on processes in reaction solutions is well known. It is enough to recall the effects it causes in the free-radical polymerization of vinyl monomers [68,84,155]. As for catalytic reactions involving fullerene, the author knows of only two: the oligomerization of terpenes [156] and the polymerization of styrene [83,84,155]. In the first case, the catalytic behavior of fullerene was suspected, while in the second, it has been convincingly proven. For a detailed description, we recall that catalysis has always been and remains the most difficult part of synthetic chemistry and is determined by a large number of thermodynamic and kinetic factors. In turn, the thermodynamics and kinetics of any elementary reaction between a pair of reactants are described by a characteristic $E(R)$ graph [157,158], the main parameters of which are shown in the inset of Figure 6. The enthalpy of the reaction, or the coupling energy between its participants $\Delta H = E_{cpl}$, determines whether the reaction will be energetically favorable. The height of the energy barrier $E_{bar} = E_{aa}$ determines the reaction rate $k_{rc}$. These two quantities are the main numerical descriptors of the reaction, and it is with them that the catalyst must work when the reaction is difficult or too slow. Obviously, the reaction is favored by $E_{cpl}$, large in absolute value and negative in sign. If the thermodynamic descriptor of the reaction under consideration does not meet this requirement, the catalyst takes over the initial stage of the reaction, forming a pair with one of the participants with a good descriptor $E_{cpl}$. A typical example is the free radical polymerization of vinyl monomers, in which free radicals act as effective catalysts of the first step of the reaction (see a detailed description in [82,84]). In turn, the height of the energy barrier $E_{bar}$ in the case of participants from covalent chemistry depends on the energy gap $E_{gap} = I_D - \varepsilon_A$, where $I_D$ and $\varepsilon_A$ are the ionization potential and the electron affinity energy of the donor and acceptor participants in the reaction, respectively [159], so that a decrease in $I_D$ or an increase in $\varepsilon_A$ is accompanied by a decrease in the value of $E_{bar}$ and a corresponding increase in the kinetic descriptor $k_{rc}$. This circumstance is one of the main ones for metal catalysts characterized by low $I$. A typical example is the kinetically hindered dimerization of fullerene C$_{60}$ [160] and its monoderivatives [155], which becomes practically barrier-free in an electric field [160] or upon the addition of butyllithium to C$_{60}$ [155].

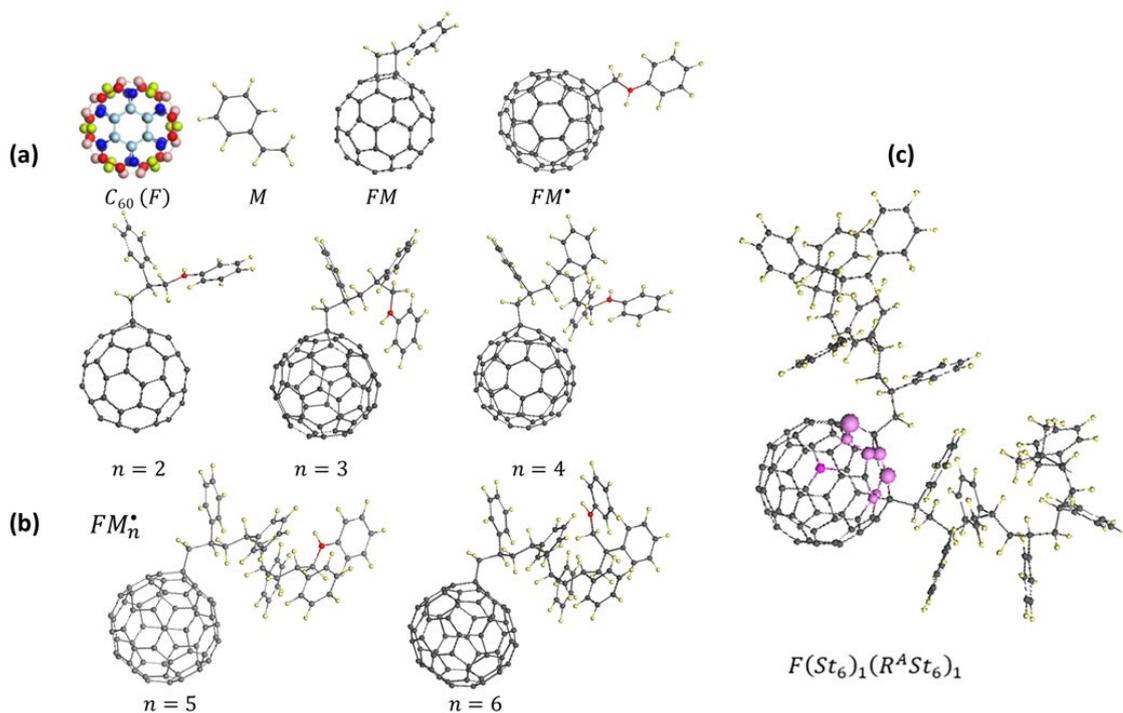

**Figure 15.** Digital twins of the virtual polymerization of styrene catalyzed with fullerene $C_{60}$. (a) Reaction participants. (b) $FM_n^\bullet$ oligomers for n from 1 till 7. (c) Two-branch element from a star-branch polymer product. UHF AM1 calculations.

A virtual quantum-chemical consideration of the thermodynamic and kinetic descriptors $E_{cpl}$ and $E_{bar}$ of elementary reactions in a reaction solution containing styrene, the free radical $AIBN$, and fullerene $C_{60}$ showed that the fastest reaction is the formation of the monomer-radical $FM^\bullet$, in which fullerene $C_{60}$ plays the role of the free radical, provided that the monomer is added to it through a single-bonded intermolecular contact [83,155] as shown in Figure 15a. As can be seen from the figure, $FM^\bullet$ confidently leads polymerization of styrene, the first members of which are shown in Figure 15b. Additionally, the formed oligomer-radical branches readily attach to the remaining free portions of the carbon frameworks of the fullerenyls generated during the reaction, forming star-shaped structures of the final polymerization products, represented in Figure 15c as an intermediate star $F(St_6)_1(R^A St_6)_1$. The resulting virtual picture is confirmed by a detailed analysis of the empirical data.

Following the path taken to establish the catalytic activity of fullerene, let us return to the oligomerization of α- and β-pinenes in the presence of fullerene $C_{60}$ and gaseous oxygen [156]. Thus, the reaction solution of interest contains a pinene, $C_{60}$ fullerene, and degassed water. As follows from the experiment, pinene oligomerization is not observed; the monomer and fullerene retain the properties of the original molecules, indicating the absence of a productive interaction between them. The situation changes after passing a stream of oxygen through this solution. The transparent free-flowing water becomes viscous. Its color caused by fullerene [101] changes noticeably, indicating the formation of fullerosils. Fast atom bombardment mass spectrometric analysis reveals the presence of terpene's dimers, trimers, and even traces of tetramers, which naturally explains the occurrence of the water viscosity, as well as fullerosils of the compositions $C_{60}O$, $C_{60}O_2$, $C_{60}O_3$. Since separately pinene and fullerene dissolved in degassed water do not lead to similar changes, the conclusion follows that these changes are due to the interaction of the components of the pinene-fullerene-oxygen triad. What happens to this triad? To answer this question, it is necessary to consider virtually a set of elementary reactions, similar to what was done in the case of vinyls' polymerization [82,84]. However, even without performing calculations, a certain prediction based on the experience of the styrene polymerization discussed above, can be made.

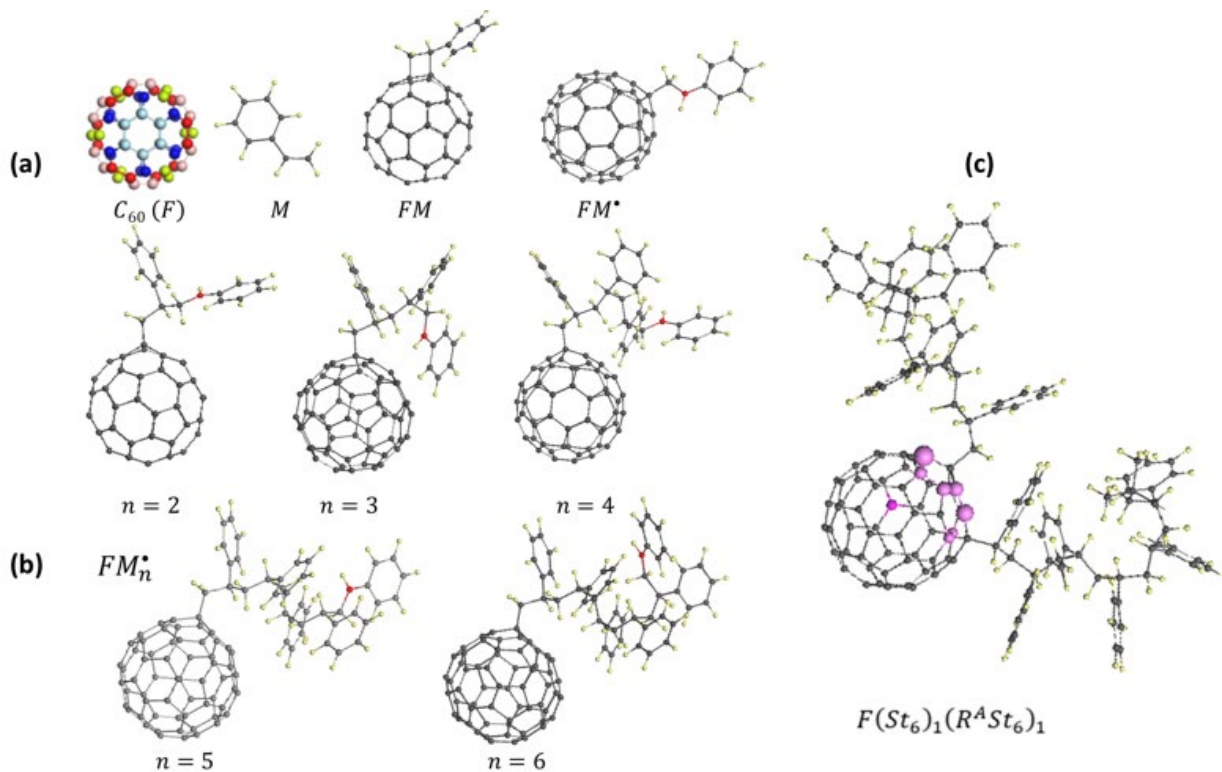

**Figure 15.** Digital twins of the styrene polymerization. (a) The content of virtual reaction solution. (b) Styrene oligomers. (c) First members of the $C_{60}$-branched star polymer formation in VRS of styrene. Size of lilac balls correspond to the ACS $N_{DA}$ values on the atoms. UHF AM1 calculations.

As follows from structure measurements [161], the particularly active sites of $\alpha$- and $\beta$-pinene molecules are connected via $sp^2 C = C$ bond of their vinyl groups of 1.34 Å in length [161]. This value is much lower than $R_{cov}^{db}$, so that odd electrons of both carbon atoms are completely paired, as a result of which $N_{DA}$ on both atoms is zero. Since the remaining carbon atoms are linked by alkane bonds and there are no other types of heteroatoms, the original pinene molecule is chemically inactive and its oligomerization, as in the case of benzene molecules [69, chapter 2], should be difficult. The same can be said about the oxidation of the molecule. In contrast, the oxidation of fullerene $C_{60}$ and the formation of the aforementioned fullerosils are quite expected [67, 68]. Moreover, each new added oxygen atom breaks one of the $sp^2 C = C$ bond leaving the paired carbon atom with high $N_{DA}$ value as a target for the next addition. If pinene is this addend, only one-bond connection between fullerene and one of its vinyl group carbons is possible which leads to generation of monomer radical $FM^\bullet$ analogous to that of styrene one discussed above. The resulting radical $FM^\bullet$ ensures the oligomerization of pinene on the fullerene body in the form of fullerosil $C_{60}M_n$. The bulkiness of the terpene molecule and the resulting steric hindrances evidently prevent the formation of long oligomers. Room temperature of the solution leads to the splitting off the resulting oligomers into the solution. Virtualization of this process is entirely feasible, just like any other involving a monomer with an alkene bond, thus opening the way to virtual fullerene carbocatalysis.

## 7. Aposteriori Reflections and Conclusive Remarks

The $C \leftrightarrow C$ bond type depicts the carbon family in which carbon-containing substance are born. Odd electrons, generated at either formation or rupture of the bonds, and spin-radical interaction between them are the main contributors to the treasure trove of special properties of these substances. This is how the credo of carbon covalent chemistry can be summarized today. Its first part reflects the trimodality of the $C \leftrightarrow C$ bonds as a set of *$sp^3 C - C$, $sp^2 C = C$,* and *$sp^1 C \equiv C$* ones, while the second refers to the peculiarities of the weak electron interaction, revealing its spin nature and transforming the classical covalent chemistry of carbon into spin covalent one. The degree of importance of the weak interaction varies for each of the three bond types, but in all cases is determined by the bonds' length. This article represents the first review of existing results in carbon chemistry, considered from the perspective of the spin nature of $C \leftrightarrow C$ bonds. Without rehashing the content of the article, we will formulate its main conclusions in the bond-length language.

*$sp^3 C - C$ bonds, known as single and alkane bonds.* The life of these bonds occurs in the covalent regime, in the absence of odd electrons and the associated spin effects, which manifest themselves only at the moment of their rupture. Using the example of the ethane molecule, the life cycle of its lone $sp^3 C - C$ bond lasts in the region of length $\Delta W_{cov}^{sg}$ = 0.607 Å that begins at $R_{eq}^{sg}$ = 1.503 Å, and ends when the bond length $l^{C-C}$ reaches $R_{cov}^{sg}$ = 2.110 Å, at which the bond starts to radicalize sharply and then goes to full rupture. As we can see, $\Delta W_{cov}^{sg}$ reveals a fairly wide range of possible bond elongation, while maintaining its spinless existence. After the bond rupture, two free radicals emerge with the summarized number of effectively unpaired electrons $N_D$ equal to 2. The equilibrium $R_{eq}^{sg}$ and critical $R_{cov}^{sg}$ within 2-3% retain the given values in almost all cases of observing these bonds in molecules not only carbonaceous, but also containing various heteroatoms. The values of $R_{eq}^{sg}$ are confirmed by numerous structural data. As for $R_{cov}^{sg}$, spin covalent chemistry of carbon assigns it the place of the reaction coordinate localizing the transition state in reactions caused by either formation or rupture of this bond.

*$sp^2 C = C$ bonds are known as double, alkene, and aromatic.* The life cycle of the lone bond of ethylene within the limits of its length $\Delta W_{cov}^{db}$ = 0.062 Å from $R_{eq}^{db}$ = 1.326 Å to $R_{cov}^{db}$ = 1.388 Å occurs in classical spinless regime, when odd electrons are tightly bound providing the $\pi$ electron contribution into the bond additionally to the $\sigma$ one. When the bond length $l^{C=C}$ exceeds $R_{cov}^{db}$, the spin covalence begins to work stimulating at first radicalization and breaking the $\pi$ constituent of the bond reaching $N_D$ = 2 e by $R_{k1}^{db}$ = 2.140 Å and accompanying further elongation of the bond by radicalization of its $\sigma$ component up to complete bond rupture and growth of the number of effectively unpaired electrons $N_D$ to 4. The $sp^2 C = C$ bond behavior above $R_{k1}^{db}$ = 2.140 Å is similar to that of the $sp^3 C - C$ one above $R_{cov}^{sg}$ = 2.110 Å.

$\Delta W_{cov}^{db}$ is the main spin-marker of each $sp^2 C = C$ bond. The value may be small and big, as well as, what is the most important, positive and negative. If it is positive, the bond is of classical spinless format and molecular compositions on their basis are spinless and non-radical. Therewith, the small marker value does not change the molecule behavior as it is in the case of benzene, for which $\Delta W_{cov}^{db} \approx 0.005$ Å. However, already in the case of naphthalene, we encounter a situation where the carbon framework of the molecule is formed by bonds of two types: short and long, with $\Delta W_{cov}^{db}$ positive for the former and negative for the latter. Thus, classical and spin covalence begin to compete, and the final result depends on the number of both bonds. The situation becomes sharp since 'the safety interval' $\Delta W_{cov}^{db}$ is not very large making up only 10% of that one for alkane bonds, thus not presenting serious obstacles to the flow of alkene bonds from the group of short bonds to the group of long bonds and vice versa. In the PAH polyacene series, as the number of benzene rings increases, the number of long bonds increases faster than short ones, which correlates with the increasing radicality of the molecules, to the point where their synthesis at ambient conditions becomes impossible. We encounter a similar situation in the case of $sp^2$ nanocarbons – fullerenes, carbon nanotubes and graphene domains. Clearly, the ratio of the numbers of negative (long bonds) and positive (or short bonds) determine the stable-radical status of these molecules. It appears that this ratio cannot be greater than one. Actually, close-to-one ratio is characteristic for stable PAHs and mentioned $sp^2$ nanocarbons. However, more exact answer to this question requires further research.

The case of negative $\Delta W_{cov}^{db}$ markers for all double bonds evidences highly reactive $sp^2$ radical species. Molecular $sp^2$ carbon of this type, neither virtual, nor real, has not been known. In contrast, it is typical case for virtual $sp^2$ silicon. Even for lone $sp^2 Si = Si$ bond of disilene, $\Delta W_{cov}^{db}$ = -0.72 Å. This value is preserved with accepted accuracy in large molecules such as aromatic hydrosilicons similar to polyacenes and $sp^2$ nanosilicons in form of fullerenes, tubes, and one-atom thick domains leaving no hope for the synthesis of the species in reality. However, there are still quite a few researchers who do not accept this verdict of spin covalent chemistry and continue to discuss, in particular, the properties of silicene as a real object analogous to graphene.

Carbon fullerenes, nanotubes, and necklaced graphene domains are stable radicals, properties of which are subordinated to spin theory of radicals. Thus, they are susceptible to chemical modification in various ways, in particular, to various derivatizations. For closed structures, such as frameworks of these substances, this property may lead to the return of spin-covalent $sp^2 C = C$ bonds into classical spinless format. The fact is that any addition act with respect to a framework atom causes an inevitable $sp^2 \rightarrow sp^3$ transformation of valence electrons hybridization of the latter. This rearrangement is accompanied with changing of spatial structure of the framework. Thus, the flat trigonal packing of $C_3$ triad atoms connected with $sp^2 C = C$ bonds is transformed into a three-dimensional tetrahedral packing of the $C_4$ tetrahedron atoms connected with $sp^3 C - C$ bonds. In the case of chemical modification of $sp^2$ nanocarbons in full, this leads to the replacement of flat benzenoid basic units with their cyclohexanoid analogs. As a result, the chemical action is accompanied by physical restructuring, causing significant mechanical stress since the $sp^2 \rightarrow sp^3$ rearrangement of the atomic system, requires an increase in space. Truncated icosahedron of $C_{60}$ fullerene, cylindrical packings of carbon nanotubes, or graphene membrane domains, resist this stress which causes shortening some of remaining chemically unmodified $sp^2 C = C$ bonds thus transforming their negative $\Delta W_{cov}^{db}$ markers into positive ones, which was discussed above. This leads to removing the radicalization of the corresponding $sp^2$ bonds, zeroing $N_{DA}$ and stopping chemical reactions in which these bonds are involved. This effect, exceptionally important for modern fullerenics and grapheneics, was discovered during virtual fluorination [67,162] and hydrogenation [67,163] of fullerene $C_{60}$, as well as hydrogenation of the graphene domain [69,164]. However, it was not taken into account for a long time until researchers encountered the paradoxical fact of a full identity of the characteristic Raman spectra of $GO$ and $rGO$ presented in Figure 16. Virtual stepwise oxidation of the graphene domain [140] revealed the impossibility of its oxidation in full due to a significant shortening of approximately 20% of the $sp^2$ $C_2$ pairs in the basal plane of original domains to 1.35 Å, which is significantly less than the critical value $R_{cov}^{db}$. The percentage is fully sufficient to explain the presence of 20–10% unoxidized carbons in the sample array, which has been repeatedly established in practice (see reviews [165-167]. This finding of shortened $sp^2 C = C$ bonds made it possible to explain the presence in the Raman spectrum of molecular $GO$ of the G band of fully symmetrical vibrations of $sp^2 C = C$ bonds next to the D band, which is caused by the vibrations of the same type but related to basic massive of $sp^3 C - C$ bonds. In turn, the presence of the D band in the Raman spectrum of $rGO$ raised a similar question, while, the presence of G was is absolutely natural. The answer concerns the $sp^2 \rightarrow sp^3$ transformation again.

As shown [165], generally, rGO consists of multilayer stacks of BSUs presented with nanosize necklaced graphene domains [37]. Nanoscopic buckling of the latter forces some atoms of adjacent layers generate the fourth bond, which is supported with large enough 'safety interval' $\Delta W_{cov}^{sg}$, thus transforming the entire set of valence bonds of these atoms into $sp^3$-type bonds abd ensuring D band emergence. The intensity and shape of this band allow to estimate the number of these domains in the BSU stack and the transverse size of the domains. The discussed characteristic doublet pattern of the Raman spectrum is widely used in practice as the main analytical evidence of the graphene nature of the material used, while not being able to separate the completely different $sp^3$ $OG$ and $sp^2$ $rGO$.

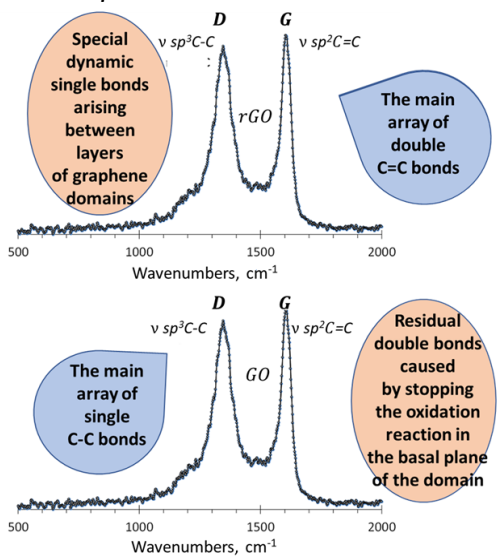

**Figure 16**. Raman spectra of graphene oxide and reduced graphene jxide

$sp^1 C \equiv C$ *bonds, known as triple or alkyne bonds.* The life cycle of a sole bond in acetylene or propyne in spinless classical format is limited with very narrow region $\Delta W_{cov}^{tr}$ = 0.043 Å from $R_{cov}^{tr}$ = 1.240 Å to $R_{eq}^{tr}$ = 1.197 Å. The marker value indicates that behavior of $sp^2$ and $sp^1$ bonds is quite similar. Convincing confirmation of this statement can be seen in the ACS $N_{DA}$ maps presented in Figure 12a for the border-regime $sp^2 C = C$ bonds of benzene and $sp^1 C \equiv C$ bonds of carbyne chains. The lone $sp^1 C \equiv C$ bond cleavage starts at the point $R_{k2}^{tr}$ = 2.100 Å, practically coinciding with $R_{cov}^{sg}$ of ethane and showing that all three lone $C \leftrightarrow C$ bonds break when their lengths reach the same value. The border-regime character of $sp^1 C \equiv C$ bonds naturally explains a comparative sparseness of population of the molecular world containing $sp^1 C \equiv C$ bonds. The bonds' chain creation is evidently impossible. The situation is saved by the possibility of the existence of alternating polyyne chains $(-C \equiv C - C \equiv C-)$. The presence of $sp^1 C \equiv C$ bonds suggests the spin-covalent nature of the doubly bonded links of these chains, however, the problem should be proven virtually.

The discussion, presented in this review in the language of valence bonds, is a discussion of the essence of covalent chemistry. This language was formed during the development and maturation of classical spinless covalent chemistry over a long historical period. Spin covalent chemistry has emerged from the vast experience of its predecessor and retains its language, imbuing it with new meanings and content. A significant advantage of the new scientific concept is the spin theory of electron-electron interactions, with particular emphasis on the interaction's weakness under certain circumstances. This theory utilizes a well-developed mathematical apparatus, allowing for the construction of transparent algorithms describing the behavior of atoms in molecules and facilitating the widespread digitalization of individual chemical processes and the construction of their virtual worlds [168].

The review mentions the first results of virtual fullerenics, virtual graphenics, virtual polymerization of vinyl monomers, virtual graphene and fullerene carbocatalysis. But the main result of spin covalent chemistry is the obtained convincing proof of the uniqueness of the carbon atom as the only pretender on the king throne of covalent chemistry, while the closest contenders from the tetrel family - silicon, germanium, and tin - are deprived of this right according to the law of spin covalent chemistry due to the deprivation of their possibility of material embodiment. As for the spin covalent chemistry of carbon, it is taking its first steps, but even these still few successes open up broad horizons for a new vision of processes controlled by spin-radical electronic interactions, such as chemical modification or derivatization of any complexity, polymerization and catalysis of molecular compounds whose electronic properties have a spin aroma.

**Acknowledgments.** This paper has been supported by the RUDN University Strategic Academic Leadership Program.